\documentclass[aps,prl,longbibliography,amsmath,amssymb,groupedaddress,superscriptaddress,floatfix,twocolumn,showkeys,nofootinbib]{revtex4-1}
    \usepackage{graphicx}
    \usepackage{dcolumn}
    \usepackage{bm}
    \usepackage{color}
    \usepackage{xcolor}
    \usepackage{multirow}
    \usepackage{ulem}
    \usepackage{siunitx}
    \DeclareSIUnit{\sqrthz}{\ensuremath{\sqrt{\text{\hertz}}}}
    \usepackage{soul}
    \usepackage{makecell}
    \usepackage{svg}
    \usepackage{latexsym}
    \usepackage{amssymb}
    \usepackage{textcomp}
    \usepackage{hyperref}
    \usepackage{physics}
    \usepackage{microtype}

    \newcommand{\Rb}{$^{87}$Rb }
    
    \usepackage{siunitx}
    \newcommand{\MHz}{\mega\hertz}
    \newcommand{\GHz}{\giga\hertz}

    \newcommand{\nm}{\nano\meter}

    \newcommand{\mm}{\milli\meter}

    \newcommand{\um}{\micro\meter}

    \DeclareSIUnit\gauss{G}
    \usepackage[T1]{fontenc}

    \makeatletter
    \renewcommand{\c@secnumdepth}{0}
    \makeatother

\hypersetup{
   colorlinks=true,
   linkcolor=blue,
   filecolor=blue, 
   citecolor=blue,
   urlcolor=blue,
}
\usepackage[doipre={DOI:~}]{uri}

\newcommand{\mbf}{\mathbf}

\newcommand{\BBR}{\Gamma_\text{BBR}}
\newcommand{\ava}{\Gamma_\text{ava}}
\newcommand{\PRyd}{P_\text{Ryd}}

\begin{document}

\title{Cryogenic suppression of correlated loss in Rydberg atom arrays}

\author{Zhenpu~Zhang}
    \thanks{These authors contributed equally}
    \affiliation{JILA, National Institute of Standards and Technology and University of Colorado, Boulder, Colorado 80309, USA}
    \affiliation{Department of Physics, University of Colorado, Boulder, Colorado 80309, USA}
\author{Brendan~P.~Marsh} 
    \thanks{These authors contributed equally}
    \affiliation{JILA, National Institute of Standards and Technology and University of Colorado, Boulder, Colorado 80309, USA}
    \affiliation{Department of Physics, University of Colorado, Boulder, Colorado 80309, USA}
\author{Ting~You~Tan} 
    \thanks{These authors contributed equally}
    \affiliation{JILA, National Institute of Standards and Technology and University of Colorado, Boulder, Colorado 80309, USA}
    \affiliation{Department of Physics, University of Colorado, Boulder, Colorado 80309, USA}
\author{Quinn~Manning}
    \affiliation{JILA, National Institute of Standards and Technology and University of Colorado, Boulder, Colorado 80309, USA}
    \affiliation{Department of Physics, University of Colorado, Boulder, Colorado 80309, USA}
\author{Daniel~H.~Slichter} 
    \thanks{Current address: Google Quantum AI, Boulder, Colorado 80301, USA}
    \affiliation{Time and Frequency Division, National Institute of Standards and Technology, Boulder, Colorado 80305, USA}
\author{Adam~M.~Kaufman}
    \affiliation{JILA, National Institute of Standards and Technology and University of Colorado, Boulder, Colorado 80309, USA}
    \affiliation{Department of Physics, University of Colorado, Boulder, Colorado 80309, USA}
\author{Cindy~A.~Regal}\email{regal@colorado.edu}
    \affiliation{JILA, National Institute of Standards and Technology and University of Colorado, Boulder, Colorado 80309, USA}
    \affiliation{Department of Physics, University of Colorado, Boulder, Colorado 80309, USA}

\date{\today}

\begin{abstract}

Blackbody radiation (BBR) poses a central challenge for quantum science with Rydberg atoms. By driving transitions to nearby opposite-parity states, BBR shortens Rydberg lifetimes and has long been suspected to trigger ``avalanche'' events known to cause correlated atom loss.
Here, we suppress avalanche decay and establish BBR as its dominant trigger using a Rydberg atom array that operates either cryogenically or at room temperature.
A $\SI{45}{\kelvin}$ radiation shield reduces ambient BBR, extending Rydberg $nS$ lifetimes for $n$ ranging from 50 to 75. BBR-induced decay rates are suppressed by \SIrange{30}{50}{\percent}, corresponding to an effective blackbody temperature $T_{\rm eff}\approx\SI{190}{\kelvin}$. This enables controlled comparisons of Rydberg-dressed atom arrays operating in different BBR environments. 
Arrays of up to 70 atoms demonstrate cryogenically suppressed avalanche rates, tracking the suppression in BBR-induced decay, for both $n=50$ and $n=60$ dressing.   
A parameter-free rate model agrees with the measured avalanche rates after accounting for frequency-dependent cryogenic shielding, photoionization-triggered avalanches, laser phase noise, BBR decay channels, and Rydberg interaction effects.
We demonstrate a quantitative description of avalanche decay and show that improved BBR shielding should further suppress correlated loss to the $T_\text{eff}=\SI{45}{\kelvin}$ achievable in the current apparatus. Our results establish a route to mitigating correlated errors that otherwise impede fault-tolerant operation in neutral-atom quantum processors.

\end{abstract}

\maketitle

\let\savedaddcontentsline\addcontentsline
\renewcommand{\addcontentsline}[3]{}

Neutral-atom arrays provide a versatile platform for quantum science~\cite{kaufman2021quantum,browaeys2020many}.
The strong, programmable interactions between Rydberg states~\cite{saffman2010quantum}, together with continuing improvements in coherent atomic control, are enabling rapid progress in quantum metrology~\cite{eckner2023realizing,bornet2023scalable,cao2024multi}, many-body quantum simulation~\cite{ebadi2021quantum,scholl2021quantum,semeghini2021probing,scholl2023erasure,chen2023continuous}, and quantum computation~\cite{graham2022multi,bluvstein2026fault,zhang2026logical}. Yet atom arrays are immersed in blackbody radiation (BBR):
Rydberg states with principal quantum number $n\gtrsim 50$ are separated by tens of gigahertz or less~\cite{li2003millimeter,mack2011measurement}, frequencies at which blackbody photons have occupation numbers in the hundreds at room temperature~\cite{gallagher1994rydberg}.
BBR-induced stimulated emission and absorption are major Rydberg decay channels at room temperature~\cite{beterov2009quasiclassical}, contributing to errors in state-of-the-art two-qubit gates~\cite{evered2026high,tsai2025benchmarking,senoo2026high}.
BBR also imposes AC Stark shifts even on low-$n$ states, a key systematic in atomic frequency standards at both microwave~\cite{Itano1982, Beloy2006, Angstmann2006} and optical frequencies~\cite{ludlow2015optical,aeppli2024clock,hassan2025cryogenic}. 

BBR is also postulated to trigger catastrophic atom loss through an avalanche decay mechanism~\cite{goldschmidt2016anomalous,zeiher2016many,boulier2017spontaneous} observed in analog quantum simulators~\cite{geim2026engineering} and in Rydberg dressing experiments that off-resonantly couple to the Rydberg state with a dressing laser~\cite{johnson2010interactions,pupillo2010strongly}.
BBR transfers atoms into nearby opposite-parity Rydberg states, termed ``contaminant'' states, whose strong dipolar interactions can shift Rydberg levels closer to or into resonance with the dressing laser~\cite{festa2022blackbody}. 
Contaminants thereby facilitate resonant Rydberg excitation in nearby atoms, locally accelerating the rate of BBR-induced contaminant production in a self-reinforcing avalanche decay process.
A single contaminant could thereby trigger rapid dephasing and correlated atom loss across a large array, limiting coherence and scalability~\cite{zeiher2016many}. 
Contaminants can also perturb Rydberg gates,
introducing correlated faults that compromise quantum error correction~\cite{bluvstein2026fault}. 

Further progress with Rydberg atom arrays therefore increasingly requires mitigating BBR, and cryogenic environments offer a direct route by suppressing it at the source.
Cryogenic neutral-atom platforms have demonstrated extended vacuum lifetimes~\cite{Schymik2021single,zhang2025lifetime} and increased defect-free array sizes~\cite{schymik2022situ, lim2026defect} enabled by cryopumping of the vacuum environment.
By suppressing ambient BBR, cryogenic operation has been reported to extend the $30S_{1/2}$ lifetime in a dilute \Rb ensemble~\cite{mack2015all} and the Rydberg $nP$ lifetimes of $^{133}$Cs atoms in a tweezer array~\cite{jin2026extended}.   
Circular Rydberg states with enhanced lifetimes have also been realized both by cryogenic operation~\cite{cantat2020long} and by engineering the electromagnetic mode density~\cite{wu2023millisecond,holzl2024long,pultinevicius2025long}.
However, to date there has not been a controlled comparison at different temperatures to confirm BBR as the microscopic trigger of Rydberg avalanches and demonstrate the suppression of correlated loss at cryogenic temperatures. 

Here, we cryogenically suppress correlated atom loss and confirm BBR as the dominant avalanche trigger using a Rydberg atom array in a controllable blackbody environment. 
Cryogenic operation extends Rydberg $nS$ lifetimes for $50\leq n\leq75$ and decreases avalanche rates compared to room-temperature operation.
Rydberg lifetimes are cryogenically extended up to \SI{45(11)}{\percent}, corresponding to suppression of BBR-induced decay rates $\Gamma_\text{BBR}$ by up to \SI{53(8)}{\percent}.
Avalanches in Rydberg-dressed atom arrays are found to occur at a rate $\ava$ scaling with total atom number $N$ up to 70.
Cryogenic operation suppresses avalanches for all $N$ and Rydberg states studied, reducing $\ava$ by up to \SI{22(2)}{\percent} in agreement with modeled rates.
We identify trap-induced photoionization as an additional avalanche trigger relevant to two-photon Rydberg coupling schemes, which can be mitigated with shallow traps, and we characterize multiple factors that influence the avalanche rate. 
Moreover, we find that avalanches can propagate across the entire atom array due to an interplay between dipolar interactions and Rydberg dressing, highlighting the importance of cryogenically suppressing their formation.
Below, we develop these results in detail, arriving at a quantitative and experimentally validated model of avalanche decay that charts a path toward improved coherence and suppressed correlated loss in Rydberg platforms.

\section{The cryogenic Rydberg atom array}

The ability to operate either cryogenically or at room temperature in the same apparatus enables direct characterization of the effects of temperature, with other variables controlled. This is an especially important capability in the often complex microwave (MW) and thermal environments experienced by Rydberg atom arrays.  
As we discuss in Sec.~\ref{sec:singlebody}, the single-atom Rydberg lifetimes sometimes deviate from their theoretical free-space lifetimes due to Purcell effects in the cryogenic enclosure. 
Dual-temperature operation largely disentangles Purcell effects from temperature effects, which could otherwise be conflated. 

\begin{figure}[!t]
    \centering
    \includegraphics[]{ 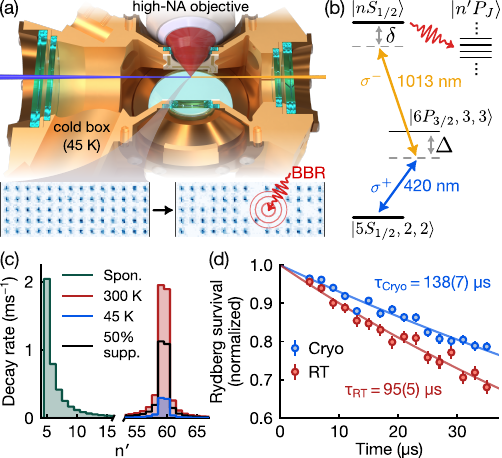}
    \caption{
    BBR suppression in a cryogenic Rydberg atom array.
    (a) Cutaway view of the \SI{45}{\kelvin} cold box (gold surfaces and transparent windows).
    Laser light (red) delivered through a room-temperature, high-NA objective forms an optical tweezer array, while two additional laser beams (blue and yellow) provide Rydberg coupling.
    Rectangular boxes show single-shot fluorescence images of a $5\times 14$ atom array before Rydberg dressing (left) and following an avalanche triggered by BBR (right). 
    (b) Rydberg coupling scheme and BBR-coupled MW transitions from $nS_{1/2}$ to nearby $n'P_J$ contaminant states.
    (c) Calculated Rydberg decay rates from $60S_{1/2}$ to $n'P_J$ at \SI{300}{\kelvin} (red) and \SI{45}{\kelvin} (blue). The green line shows the temperature-independent spontaneous emission rates.
    As an illustrative reference, the black line models the reduced BBR decay rate corresponding to a simple \SI{50}{\percent} cryogenic suppression estimate based on the cold box geometry. 
    (d) Measured cryogenic enhancement of Rydberg lifetime. Exponential fits show the $60S_{1/2}$ lifetime measured at room temperature ($\tau_\text{RT}$) and cryogenic conditions ($\tau_\text{Cryo}$).   
    Error bars denote \SI{68}{\percent} confidence intervals throughout this manuscript. 
    }
    \label{fig1}
\end{figure}

Our apparatus, details of which are provided in Ref.~\cite{zhang2025lifetime}, operates at either \SI{45}{K} or \SI{300}{K}. The cryogenic enclosure for the \Rb atom array is comprised of a radiation shield, which we refer to as the ``cold box,'' and cold windows that are designed to combine high-numerical-aperture (NA) optical access and cryogenic operation [Fig.~\ref{fig1}(a)], rather than to shape the electromagnetic density of states as in circular Rydberg apparatuses~\cite{cantat2020long,wu2023millisecond,holzl2024long,pultinevicius2025long}. An optical tweezer array is formed using an in-vacuum, high-NA objective with $>$\SI{400}{\um} diameter field of view, enabling studies of homogeneous atom arrays. The objective is located directly outside the cold box and remains at room temperature during cryogenic operation, preserving alignment. The atom array is fully surrounded by cryogenic surfaces, but room-temperature BBR still passes through the cold windows; future designs will employ optical apertures coated with optically transparent, electrically conductive indium tin oxide (ITO) or ZnO~\cite{wang2025can} to improve BBR shielding. The cold windows subtend approximately \SI{50}{\percent} of the solid angle to the atom array, providing a simplified estimate of \SI{50}{\percent} shielding from room-temperature BBR under cryogenic operation. We refine this estimate in Sec.~\ref{sec:singlebody} using the measured Rydberg lifetimes.

Atoms are loaded from a three-dimensional magneto-optical trap into an optical tweezer array generated using a spatial light modulator.  
We use $\Lambda$-enhanced gray molasses ($\Lambda$GM)~\cite{brown2019gray} on the D1 transition to achieve single-atom filling rates in excess of \SI{80}{\percent}. 
A crossed acousto-optic deflector rearranges the atoms into ordered arrays.
Rydberg dressing experiments begin with an average of 80 atoms loaded into a $5\times 20$ array, and finite-field $\Lambda$GM cooling~\cite{qiu2026loading} at a trap depth of $\SI{0.88}{\milli\kelvin}$ and a field of $\SI{5.3}{\gauss}$ yields an atomic temperature of \SI{12.0(2)}{\micro\kelvin}.
For single-atom lifetime measurements, atoms are instead loaded stochastically into a $1\times4$ array with \SI{33}{\micro\meter} spacing, which is large enough to ensure that atom-atom interaction effects can be neglected. 
The same finite-field $\Lambda$GM cooling sequence is applied, followed by a reduction in tweezer depth to \SI{20}{\micro\kelvin}, yielding an atomic temperature of \SI{0.93(6)}{\micro\kelvin}.

All reported measurements use a \SI{3.5}{\gauss} magnetic field during Rydberg excitation, applied parallel to the Rydberg beams. 
Atoms are optically pumped to $\ket{g}\equiv\ket{5S_{1/2},F=2,m_F=+2}$ and coupled to Rydberg states $\ket{r}\equiv\ket{nS_{1/2},m_J=+1/2}$ using counter-propagating, circularly polarized \SI{420}{\nm} and \SI{1013}{\nm} beams [Fig.~\ref{fig1}(a),(b)]. 
The \SI{420}{\nano\meter} Rydberg beams are red-detuned from the excited state $\ket{6P_{3/2},F=3,m_F=+3}$ with $\Delta=-2\pi\times\SI{1.5}{\GHz}$. The two-photon Rydberg coupling strength is $\Omega \leq 2\pi \times \SI{3}{\mega\hertz}$.
Details of cooling, state preparation, and the excitation sequences are provided in Appendix~\ref{supp:atom-prep-ryd-coupling}.

\section{Extended Rydberg lifetimes}\label{sec:singlebody}

We now demonstrate cryogenic enhancement of single-atom Rydberg lifetimes. Rydberg states decay at a total rate $\Gamma=\Gamma_\text{sp}+\BBR$, where $\Gamma_\text{sp}$ is the spontaneous emission rate and $\BBR$ is the rate of stimulated emission and absorption of blackbody photons. The BBR-induced component accounts for over \SI{50}{\percent} of the total decay rate for \Rb $nS$ states with $n\geq 48$, and accounts for even more for states with orbital angular momentum $L>0$~\cite{vsibalic2017arc}. 

The ambient MW environment sets $\BBR$ through the mean occupation number $\bar n$ of blackbody photons at the Rydberg transition frequencies. The BBR-induced decay rate for a Rydberg state with quantum numbers $(n,L,J,m_J)$, with total angular momentum $J$ and quantization-axis projection $m_J$, is~\cite{beterov2009quasiclassical,pultinevicius2025long} 
\begin{equation}
\begin{split}\label{eq:BBR}
    \BBR = (2J+1&)  \sum_{\substack{n'L'\\J'm'_J}}\sum_{q\in\{-1,0,1\}}\,
    \begin{pmatrix} J' & 1 & J \\ -m_J' & q & m_J \end{pmatrix}^{2} \\
    & \times A(nLJ\to n'L'J')\,
    F_P(\omega,q)\,
    \bar{n}(\omega) .
\end{split}
\end{equation}
All dipole-coupled states $(n',L',J',m'_J)$ and photon polarizations $q$ are summed over. The squared term is the Wigner 3-j symbol and $A(nLJ\to n'L'J')$ is the Einstein $A$ coefficient. The Purcell factor $F_P(\omega,q)$ modifies the local density of states and can depend on both the polarization $q$ and transition frequency $\omega$ given the complicated boundary conditions imposed by the cold box. 
In thermodynamic equilibrium, Bose--Einstein statistics gives $\bar n(\omega,T) = [\exp(\hbar \omega/k_BT)-1]^{-1}$, where $\omega$ is the photon frequency, $k_B$ is the Boltzmann constant, $\hbar$ is the reduced Planck constant, and $T$ is the temperature. 
For the dominant BBR-induced Rydberg transitions we study, $\bar n(\omega,T)\approx k_BT/(\hbar \omega)$ for $T\gg\hbar\omega/k_B\approx\SI{1}{\kelvin}$.
Cryogenic operation would then be expected to decrease $\BBR$ approximately linearly with the temperature of the MW environment.   

\begin{figure}[t]
    \centering
    \includegraphics{ 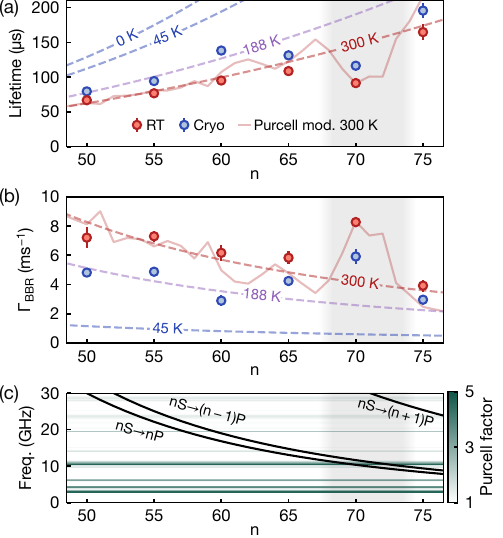}
    \caption{Cryogenic enhancement of Rydberg lifetimes and suppression of BBR-induced decay across $50\leq n\leq75$.  
    (a) 
    Measured $nS_{1/2}$ lifetimes under room-temperature and cryogenic operation.
    Dashed curves show free-space decay rates assuming isotropic BBR at the indicated temperatures~\cite{vsibalic2017arc}.
    The solid curve shows the room-temperature prediction including Purcell modifications from finite-element simulations (Appendix~\ref{supp:COMSOL}) of the cold box, with no adjustable parameters.
    (b) 
    BBR-induced decay rates inferred from the measured lifetimes in (a), found by subtracting the spontaneous emission rate $\Gamma_\text{sp}$ from the total decay rates.
    Dashed curves show the corresponding free-space predictions.
    The solid curve corresponds to the same Purcell-modified estimates above.
    (c)
    Calculated microwave transition frequencies from $nS$ to nearby $n'P$ states (black) together with the calculated polarization-averaged Purcell factor of the cold box (green). 
    The shaded region marks the range of $n$ where Purcell enhancements significantly modify the transition rates, yielding enhanced $\BBR$ near $n=70$.
    }
    \label{fig2:direct_lifetime}
\end{figure}

Measured Rydberg lifetimes for $^{87}$Rb $nS_{1/2}$ states for $n$ ranging between 50 and 75 are presented in Fig.~\ref{fig2:direct_lifetime}(a), demonstrating a cryogenic enhancement over room-temperature values in each case. To measure lifetimes, we apply a two-photon $\pi$ pulse to excite the Rydberg state, wait a variable amount of time, then measure the Rydberg survival probability using a second $\pi$ pulse to return the atom to the ground state for detection~\cite{scholl2023erasure,jin2026extended}. Representative Rydberg survival curves for room-temperature and cryogenic operation are shown in Fig.~\ref{fig1}(d). The exact timing sequence for the measurement is designed to eliminate systematic biases arising from Rydberg decay products, with details provided in Appendix~\ref{supp:lifetime-measurement}. Figure~\ref{fig2:direct_lifetime}(b) shows the inferred $\BBR$ after inverting the measured lifetimes to obtain decay rates, then subtracting the known spontaneous emission rates~\cite{vsibalic2017arc}. We observe over \SI{50}{\percent} reduction in $\BBR$ for $60S_{1/2}$; other states show similar or smaller reductions. Thus, cryogenic operation provides a controlled and variable tuning knob for altering the BBR environment.  

The degree of cryogenic BBR suppression is determined by how well room-temperature BBR is shielded by the cold box.
Under cryogenic operation with the current apparatus, room-temperature BBR transmitted through the cold windows mixes in some proportion with \SI{45}{\kelvin} radiation. Approximating the MW environment using a single effective temperature, we find that $T_\mathrm{eff}=\SI{188(8)}{\kelvin}$ matches the data well for $n\leq 60$, corresponding to $\SI{44(3)}{\percent}$ shielding from room-temperature BBR. 
While $T_\mathrm{eff}$ is a useful way to quantify the net BBR reduction, we show in Sec.~\ref{sec:ava} that the degree of BBR shielding likely depends on $\omega$, such that $\bar n(\omega)$ cannot be described by Bose--Einstein statistics at any single temperature. 
Conductive coatings~\cite{wang2025can} on the cold windows should fully equilibrate the MW environment to $\SI{45}{\kelvin}$.

The measured Rydberg lifetimes for $n\geq 65$ are modified by Purcell enhancement of the BBR transition rates. 
Finite-element simulations in Appendix~\ref{supp:COMSOL} suggest the cold box hosts MW resonances with quality factors up to 50, particularly in the $3$ to $\SI{15}{\GHz}$ band, leading to Purcell factors $F_P\lesssim 5$ [Fig.~\ref{fig2:direct_lifetime}(c)]. 
We estimate $F_P(\omega,q)$ for each polarization and for frequencies up to $\SI{30}{\GHz}$ using finite-element simulation, with the polarization-averaged results shown in Fig.~\ref{fig2:direct_lifetime}(c). These Purcell effects in the MW regime have negligible effect on $\Gamma_\text{sp}$, which is dominated by optical transitions to short-lived, low-$n$ excited states. 

Accounting for Purcell effects reproduces the enhanced $\BBR$ observed for $n=70$. The remaining discrepancies for $n=65$ and $n=75$ likely reflect imperfect modeling of the cold box, whose exact details are hard to capture in finite-element simulations. Ambient frequency-dependent MW noise may also contribute to the discrepancy; we find that our complementary metal-oxide-semiconductor (CMOS) camera used for atom imaging emits broadband MW radiation that reduces Rydberg lifetimes for $n\geq 65$. We build a camera shield made from MW-absorbing material and copper to eliminate this noise source (Appendix~\ref{supp:MWnoise}) but other sources could remain.  
To minimize complications from Purcell effects and MW noise in our analysis, we focus on the $50S_{1/2}$ and $60S_{1/2}$ states for investigating avalanches in Rydberg-dressed atom arrays.   

\section{Suppression of avalanche decay}
\label{sec:ava}

\begin{figure*}[!th]
    \centering
    \includegraphics[]{ 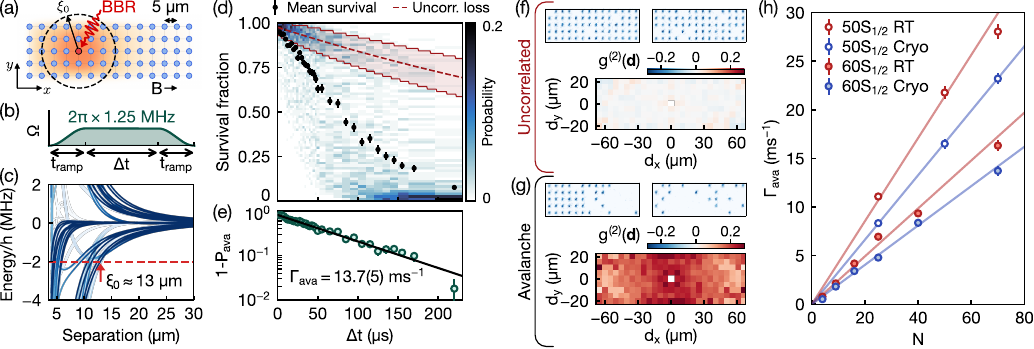}
    \caption{
    Detection and cryogenic suppression of avalanches during Rydberg dressing.
    (a) Rydberg-dressed atoms (blue) are arranged in a $5\times14$ array with \SI{5}{\micro\meter} spacing. A single BBR-induced transition produces a contaminant Rydberg atom (red), which facilitates further Rydberg excitation and contaminant generation within a radius $\xi_0$~\cite{festa2022blackbody}.   
    (b) Ramp schedule for the Rydberg coupling strength. We use cosine ramp profiles with ramp time $t_\text{ramp}=\SI{6}{\micro\second}$.
    (c) Dipolar potentials for $\ket{60S_{1/2}}\ket{n'P_J}$ pair states, including $n'=59,60,61$ and all fine structure states, shaded by their overlap with the non-interacting pair state at infinite separation. 
    Their intersection with the dressing detuning $\delta$ defines the corresponding facilitation range $\xi_0\simeq\SI{13}{\micro\meter}$ (Appendix~\ref{supp:dipolar}).
    (d) Survival distribution of the $60S_{1/2}$ Rydberg-dressed array versus $\Delta t$ under cryogenic operation. 
    Black points show the mean survival; the red dashed line and shaded region indicate the expectation and \SI{95}{\percent} binomial confidence interval for uncorrelated single-body loss.
    (e) Shots with survival below the confidence interval are classified as avalanche events. 
    The no-avalanche probability $1-P_\text{ava}$ decays exponentially (solid line), yielding a fitted avalanche rate $\ava \equiv \tau_\text{ava}^{-1}$.
    (f),(g) Representative single-shot survival patterns and spatial correlation functions $g^{(2)}(\mathbf{d})$ for shots characterized as (f) uncorrelated loss and (g) avalanche decay, both at fixed $\Delta t=\SI{45}{\micro\second}$.
    (h) Avalanche rate $\ava$ versus atom number $N$ for $50S_{1/2}$ and $60S_{1/2}$ dressing for either room-temperature (RT) or cryogenic operation. Solid lines show zero-offset linear fits.
    }
    \label{fig3:avalanche}
\end{figure*}

BBR-induced decay has long been suspected to be the microscopic trigger for avalanche decay in Rydberg-dressed ensembles~\cite{goldschmidt2016anomalous,zeiher2016many,boulier2017spontaneous,festa2022blackbody}. Even a single decayed atom can set off a chain reaction of correlated atom loss, resulting in a predicted avalanche rate $\ava\propto N \BBR$, where $N$ is the total number of atoms in the ensemble. 
We now confirm BBR's role as a trigger, identify additional avalanche triggers, and achieve quantitative agreement with predicted avalanche rates as a function of $N$, Rydberg-dressing fraction, and temperature.  

Avalanche decay is understood to originate from BBR-induced transitions to opposite-parity, contaminant Rydberg states.
Contaminants induce dipolar interactions that facilitate resonant Rydberg excitation in the remaining dressed atoms, accelerating BBR-induced decay and generating further contaminants within a characteristic range $\xi_0$~\cite{festa2022blackbody}.
This gives rise to locally correlated loss in the ensemble [Fig.~\ref{fig3:avalanche}(g)]. Existing mitigation strategies suppress contaminant buildup through stroboscopic dressing~\cite{hines2023spin,weckesser2025realization} or active autoionization~\cite{cao2025autoionization}, but require additional temporal or optical control. Here, contaminant production is instead suppressed at its source through cryogenic suppression of $\BBR$.

We perform Rydberg dressing in square-lattice arrays of up to $5\times14$ atoms with \SI{5}{\micro\meter} spacing. The long axis of the array is aligned with the Rydberg-beam propagation and quantization-field directions [Fig.~\ref{fig3:avalanche}(a)]. Tweezer traps are reduced to $\SI{9}{\micro\kelvin}$ depth to suppress photoionization effects, which we discuss later, but remain on so that dressed atoms remain trapped. Ground state atoms $\ket{g}$ are coupled to Rydberg states $\ket{r}$, where we consider both $n=50$ and $n=60$ in separate experiments. The Rydberg coupling is ramped on over $t_{\rm ramp}=\SI{6}{\micro\second}$ using a cosine profile to a maximum Rabi frequency $\Omega$, held static for a variable amount of time $\Delta t$, then ramped back down to zero in a time-reversed fashion [Fig.~\ref{fig3:avalanche}(b)]. Prior to the first BBR-induced decay, this realizes the dressing Hamiltonian 
\begin{equation}\label{eq:dressingHam}
    \frac{\hat H}{\hbar} = -\delta\sum_{i=1}^N\hat{n}_i+\frac{\Omega(t)}{2}\sum_{i=1}^N \big(\dyad{r}{g}_i + \dyad{g}{r}_i \big)+\sum_{i< j}^N V_{ij}\hat{n}_i\hat{n}_j,
\end{equation}
where $i$ and $j$ index over the atoms, $\hat{n}_i=\dyad{r}{r}_i$, and $\delta=-2\pi\times\SI{2}{\MHz}$ is the detuning from the Rydberg state. Rydberg-Rydberg interactions are described by $V_{ij}=-2\pi C_6/|\mbf{r}_i-\mbf{r}_j|^6$, where $\mbf{r}_i$ are the positions of the atoms and the dispersion coefficients are $C_6\approx\SI{-15}{\GHz\, \micro\meter^6}$ for $50S_{1/2}$ and $C_6\approx\SI{-139}{\GHz\, \micro\meter^6}$ for $60S_{1/2}$. Our maximum Rabi frequency $\Omega = 2\pi\times \SI{1.25}{\MHz}$ puts the system near the strong-dressing regime and yields a nominal Rydberg dressing fraction $P_\text{Ryd}\approx 0.08$ before considering interaction effects. 
BBR-induced decay to an $n'P$ contaminant state $\ket{r'}$ triggers spin-exchange  interactions described by the dipole-dipole Hamiltonian 
\begin{equation}\label{eq:Hdd}
    \frac{\hat{H}_\text{dd}}{\hbar}= \sum_{(i,j)}J_{ij}\big(\dyad{r}{r'}_i\dyad{r'}{r}_j + \text{H.c.}\big), 
\end{equation}
where $J_{ij}= C_3(\theta)/|\mbf{r}_i-\mbf{r}_j|^3$ depends on the angle $\theta$ between the quantization axis and the interatomic separation vector. Nearest-neighbor spin-exchange rates are estimated to 
reach up to tens of MHz at our $\SI{5}{\micro\meter}$ spacing. These spin-exchange interactions, together with higher-order dipolar interactions in the Rydberg manifold, modify the many-body excitation spectrum and can facilitate Rydberg excitation in dressed atoms, as discussed in Sec.~\ref{sec:dynamics}. 

We identify avalanche events through atom loss measured after each Rydberg-dressing sequence. 
In the absence of BBR, the atoms are dressed approximately adiabatically given $t_\text{ramp}\gg\Omega/|\delta|^2$. Atoms remain trapped during dressing due to their large coherent ground-state admixture, then return to the ground state after dressing for detection. 
Spontaneous emission results in dressed atoms decaying to the ground state manifold, remaining trapped with high probability. We use continuous optical pumping and repumping to return these atoms to $\ket{g}$, where they re-enter the dressed state. Repeated scattering, however, induces recoil heating and eventual loss, yielding an independently characterized background loss rate as described in Appendix~\ref{supp:DressParams}. 
Avalanches lead to excess loss beyond this background. Contaminant states and the resonantly excited $nS$ states they facilitate are both weakly anti-trapped by the tweezer and are photoionized by a MW pulse directly after dressing, ensuring their removal before imaging. The avalanche rate scales with $N$, and thus avalanche decay dominates the atom loss signal in large atom arrays.   

\begin{figure*}
    \centering
    \includegraphics[]{ 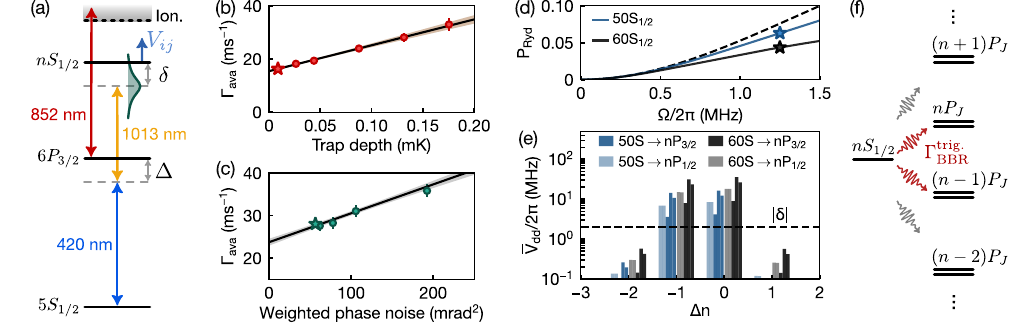}
    \caption{
    Systematic effects on avalanche rates.
    (a) Level diagram. 
    The Rydberg-coupling lasers (blue and yellow) have a phase-noise spectral density (green profile) that is nonzero at the frequency corresponding to the two-photon detuning $\delta$, which incoherently excites the Rydberg state and increases the BBR-induced decay rate during dressing. 
    The \SI{852}{\nano\meter} tweezer light excites population in $6P_{3/2}$ to above the ionization threshold (dashed line). This yields a small photoionization rate during dressing which is suppressed at a large excited-state detuning $\Delta$.  
    The arrow labeled $V_{ij}$ represents energy shifts from Rydberg-Rydberg interactions [Eq.~\eqref{eq:dressingHam}], modifying the dressing fraction.
    (b) Avalanche rate $\ava$ versus tweezer depth for a $5\times14$ array dressed with the $60S_{1/2}$ state at room temperature. Here and in (c), the black line shows a linear fit with uncertainty shown by the shaded region. Stars mark nominal dressing conditions through the figure.
    (c) $\ava$ versus integrated phase noise for a $5\times14$ array dressed with $50S_{1/2}$ state under cryogenic operation.
    The weighted phase noise is obtained by integrating the measured phase-noise spectral density against the measured Lorentzian response function (Appendix~\ref{sec:phasenoise}). 
    (d) Simulated Rydberg dressing fraction per atom $\PRyd$, including Rydberg-Rydberg interactions, versus two-photon Rabi frequency $\Omega$ (Appendix~\ref{sec:dressingfraction}). 
    Solid curves show $\PRyd$ for $50S_{1/2}$ (blue) and $60S_{1/2}$ (black); the dashed curve shows the single-atom $\PRyd$.
    (e) Calculated dipolar interaction strengths between $nS_{1/2}$ and contaminant states $(n+\Delta n)P_J$, averaged over the four nearest-neighbor directions at a separation of \SI{5}{\micro\meter}.
    The dashed line marks $|\delta|/2\pi=\SI{2}{\mega\hertz}$. Different bars correspond to different $m_J$ states in the $nP$ manifolds.
    (f) Schematic of BBR-induced transitions from $nS_{1/2}$ to nearby $n'P_J$ manifolds. The total transition rate to the strongly interacting $nP_J$ and $(n-1)P_J$ states (red) defines the avalanche-triggering decay rate $\BBR^\text{trig.}$. 
    }
    \label{fig:systematic}
\end{figure*}

We use atom loss to identify avalanches as follows. 
Figure~\ref{fig3:avalanche}(d) shows the distribution of atom loss versus dressing time $\Delta t$ for a $5\times 14$ array of Rydberg-dressed atoms under cryogenic operation. The average survival probability drops more rapidly than for a single dressed atom and the decay curve is non-exponential, as expected for rapid avalanche decay following a BBR trigger. 
However, more information is revealed by inspecting not just the average survival probability but the full distribution of surviving atoms across experimental shots. 
Independent atom loss at each site would result in a binomial distribution of surviving atoms for each $\Delta t$. We measure the loss rate for a single dressed atom separately and plot the $\SI{95}{\percent}$ binomial confidence interval in red to indicate the expected survival distribution for uncorrelated loss. The measured atom-loss distribution for the $5\times 14$ array is initially consistent with uncorrelated loss, but diverges from it with increasing $\Delta t$, indicating the onset of correlated loss. Experimental shots with atom loss exceeding the $\SI{95}{\percent}$ confidence interval for independent loss are classified as avalanche events. Avalanche decay is found to occur at a constant rate $\ava$ [Fig.~\ref{fig3:avalanche}(e)]. This rate shows little dependence on the precise choice of confidence interval (Appendix~\ref{supp:AvaDataAna}). 

Strong correlations in atom loss are detected among the experimental shots identified with avalanches. This is quantified by the spatial correlation function, which for defect-free arrays takes the form
\begin{equation}\label{eq:g2}
    g^{(2)}(\mbf{d}) = \frac{1}{\mathcal{N}}\sum_{\substack{i,j \\ \mathbf r_i-\mathbf r_j=\mathbf d}}\big\langle \big(s_i - \langle s_i \rangle\big)\big(s_j - \langle s_j \rangle\big) \big\rangle,
\end{equation}
where $s_i=1$ if an atom is detected at site $i$ and $s_i=-1$ otherwise. Angle brackets denote averages over experimental shots.
$\mathcal{N}$ is the number of atom pairs in the array with separation vector $\mbf{d}$, bounding $g^{(2)}(\mbf{d})$ between $-1$ and $1$. We consider only inter-atom correlations $|\mbf{d}|>0$. We permit up to $\SI{4}{\percent}$ defect fraction in rearranged atom arrays; this minimally affects avalanche rates but requires a more general $g^{(2)}$ formula presented in Appendix~\ref{supp:AvaDataAna}.
Shots categorized as avalanche events show strong positive $g^{(2)}(\mbf{d})$ [Fig.~\ref{fig3:avalanche}(g)], while those categorized as no-avalanche show negligible correlations [Fig.~\ref{fig3:avalanche}(f)]. This demonstrates how atom loss effectively separates avalanche-induced correlated loss from background, uncorrelated loss.

Cryogenic suppression of avalanche decay is demonstrated in Figure~\ref{fig3:avalanche}(h). We measure $\ava$ under room-temperature and cryogenic operation for both the $50S_{1/2}$ and $60S_{1/2}$ dressed states and for a variety of array sizes with total atom numbers $N$ up to 70. In all cases, $\ava$ is reduced under cryogenic operation and we observe the expected linear dependence of $\ava$ on $N$. Cryogenic operation reduces $\ava$ by $\SI{22(2)}{\percent}$ for $n=50$ and $\SI{18(2)}{\percent}$ for $n=60$.
These results directly link BBR to avalanche decay and establish that cryogenic operation can be used to suppress correlated loss.

To establish a quantitative connection between $\ava$ and $\BBR$ we account for four effects (\textit{i}--\textit{iv}) that modify $\ava$. These effects are presented in Figure~\ref{fig:systematic} and are discussed below:

\textit{(i) Trap-induced photoionization:} The \SI{852}{\nano\meter} tweezer light can photoionize atoms in the $6P_{3/2}$ excited state, producing ions with strong electric fields. Ions within a few tens of micrometers of an atom can produce megahertz-scale shifts in the Rydberg levels~\cite{engel2018observation}. Ions can therefore facilitate resonant Rydberg excitation and trigger avalanches in a similar manner to contaminant Rydberg states.
$\Gamma_\text{ava}$ is found to increase linearly with tweezer depth [Fig.~\ref{fig:systematic}(b)], consistent with the independently measured photoionization rate (Appendix~\ref{sec:ionization}). Since a single ionized atom can trigger an avalanche, we model the photoionization contribution to $\ava$ as $NP_\text{E}\Gamma_{\rm ion}$ given an excited-state photoionization rate $\Gamma_\text{ion}$ and population $P_\text{E}$. 
We quantify this contribution from the fitted linear dependence of $\ava$ on tweezer depth. The difference in $\ava$ between the rate at our nominal $\SI{9}{\micro\kelvin}$ operating depth and the extrapolated zero-depth value gives an estimated $P_\text{E}\Gamma_{\rm ion}=\SI{12(1)}{\per\second}$.
This accounts for about $\SI{5.1(6)}{\percent}$ of the total avalanche rate and is close to that predicted from theoretical estimates (Appendix~\ref{sec:ionization}).  

\textit{(ii) Rydberg-laser phase noise:} Phase noise around the frequency of the dressing detuning $\delta$ incoherently populates the Rydberg state~\cite{festa2022blackbody}, increasing the probability of BBR decay by adding a contribution $P_\text{PN}$ to the Rydberg population. We characterize this decay channel by deliberately injecting phase noise to the dressing laser and measuring the change in $\Gamma_\text{ava}$ to determine a phase-noise susceptibility [Fig.~\ref{fig:systematic}(c) and Appendix~\ref{sec:phasenoise}]. Extrapolating back to zero phase noise provides an estimation of the phase-noise contribution $P_\text{PN}\approx0.01$, typically accounting for $\SIrange{15}{20}{\percent}$ of $\Gamma_\text{ava}$.

\textit{(iii) Many-body effects in Rydberg excitation:} Van der Waals interactions between $nS_{1/2}$ states, as described in Eq.~\eqref{eq:dressingHam}, suppress the dressing fraction $\PRyd$ compared to that for a single dressed atom. We estimate $\PRyd$ in large atom arrays through exact diagonalization of $\hat{H}$ in a simplified geometry (Appendix~\ref{sec:dressingfraction}), which predicts $\PRyd= 0.063$ for $50S_{1/2}$ dressing and $\PRyd= 0.044$ for $60S_{1/2}$ dressing, reduced from the $\PRyd= 0.076$ predicted for a single dressed atom [Fig.~\ref{fig:systematic}(d)].

\textit{(iv) Weakly interacting contaminants:} Not all contaminant states trigger avalanches. 
Only when $nS_{1/2}$ dressed atoms decay to $nP$ or $(n-1)P$ states do the dipole-dipole interaction strengths at our $\SI{5}{\micro\meter}$ lattice spacing become comparable to or exceed $|\delta|$, as required for resonant facilitation and triggering avalanches [Fig.~\ref{fig:systematic}(e)]. We define the total BBR-induced decay rate to these avalanche-triggering states as $\BBR^\text{trig.}$, corresponding to Eq.~\eqref{eq:BBR} but with the sum over states limited to the $nP$ and $(n-1)P$ states [Fig.~\ref{fig:systematic}(f)]. 
Single-atom lifetime measurements determine the total $\BBR$ but do not resolve individual transition rates. 
At room temperature, the microwave photon occupation follows a thermal distribution at \SI{300}{\kelvin}, allowing us to calculate $\BBR^\text{trig.}\approx 0.7\,\BBR$ 
for both $n=50$ and $n=60$ dressing at $\delta=-2\pi\times\SI{2}{\MHz}$ (Appendix~\ref{sec:seedingefficiency}).
Under cryogenic conditions, $\BBR^\text{trig.}$ depends on the degree to which room-temperature BBR is suppressed at the specific $nP$ and $(n-1)P$ transition frequencies, which is not known. We therefore use the general bound $\BBR^\text{trig.}<\BBR$ for cryogenic operation.

\begin{figure}[!t]
    \centering
    \includegraphics[]{ 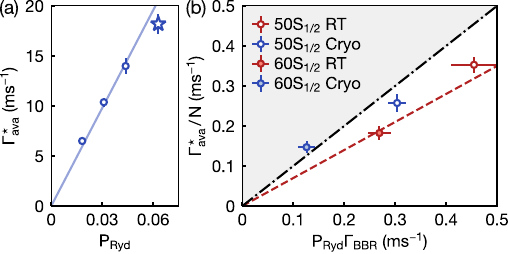}
    \caption{
    The systematics-corrected avalanche rate $\ava^{\star}$ and its dependencies. 
    (a) The Rydberg-dressing fraction $\PRyd$ is varied by reducing $\Omega$ from its nominal value of $2\pi\times\SI{1.25}{\mega\hertz}$, which reduces $\PRyd$ as shown in Fig.~\ref{fig:systematic}(d). The solid line is a zero-offset linear fit and the star marks nominal operating conditions.
    (b) The per-atom rate $\ava^\star/N$ is obtained from the fitted slope in Fig.~\ref{fig3:avalanche}(h) and plotted against $\PRyd\BBR$, derived from the single-atom lifetime measurements in Fig.~\ref{fig2:direct_lifetime}.  
    The black dash-dotted line represents the upper bound $\BBR^\text{trig.}=\BBR$ while Eq.~\eqref{eq:ava} predicts all data should fall below the shaded region. The red dashed line shows the room-temperature estimate $\BBR^\text{trig.}\approx0.7\,\BBR$.
    }
    \label{fig:avaStar}
\end{figure}

Taking into account the above avalanche triggers and effects yields a refined model for the avalanche rate,
\begin{equation}
\Gamma_{\rm ava} = N(P_{\rm Ryd}+P_{\rm PN})\BBR^\text{trig.} + NP_\text{E}\Gamma_{\rm ion}.
\label{eq:ava}
\end{equation}
We validate this model by
independently characterizing each term above and comparing the predicted $\ava$ from Eq.~\eqref{eq:ava} to the measured one.  
The photoionization contribution $N P_\text{E}\Gamma_\text{ion}$, phase-noise contribution $NP_\text{PN}\BBR^\text{trig.}$, and many-body effects on $\PRyd$ are estimated using the methods described above in (\textit{i}--\textit{iv}).
For room-temperature experiments, $\BBR^\text{trig.}\approx 0.7\,\BBR$ is obtained from the measured $\BBR$ reported in Fig.~\ref{fig2:direct_lifetime}(b), while under cryogenic conditions we can only bound $\BBR^\text{trig.}<\BBR$ from the measured $\BBR$. 
To assist in the comparison between the predicted and measured $\ava$, we define a systematics-corrected avalanche rate
\begin{equation}\label{eq:avaCorr}
    \ava^{\star} = \ava - N P_{\rm PN}\BBR^\text{trig.} - NP_\text{E}\Gamma_{\rm ion}.
\end{equation}
This gives $\ava^{\star}= N \PRyd \BBR^\text{trig.}$ as a testable hypothesis that links our measured avalanche rates in dressed atom arrays to the measured Rydberg lifetimes. Figure~\ref{fig:avaStar}(a) demonstrates the predicted linear scaling of $\ava^\star$ with $\PRyd$. 
Figure~\ref{fig:avaStar}(b) shows that the measured $\ava$ at room temperature are in agreement with the rate predicted by Eq.~\eqref{eq:ava}. The measured $\ava$ under cryogenic operation are consistent with the bound $\BBR^\text{trig.}<\BBR$.   
This agreement demonstrates the quantitative link between BBR and avalanche decay.

In addition, Fig.~\ref{fig:avaStar}(b) indicates that $\BBR^\text{trig.}$ is closer to $\BBR$ under cryogenic operation than it is at room temperature. 
This indicates that, in the current apparatus, BBR at the $nP$ and $(n-1)P$ transition frequencies is cryogenically suppressed less than BBR at the higher-frequency transitions to other $n'P$ states. 
The cryogenic BBR environment therefore cannot be fully described by a single effective BBR temperature $T_\text{eff}$ for the current apparatus. Future studies with MW environments that equilibrate to a uniform $T_\text{eff}$ under cryogenic operation should achieve the same degree of suppression in $\BBR$ and $\ava^{\star}$. 

\begin{figure*}[th]
    \centering
    \includegraphics[width=\linewidth]{ 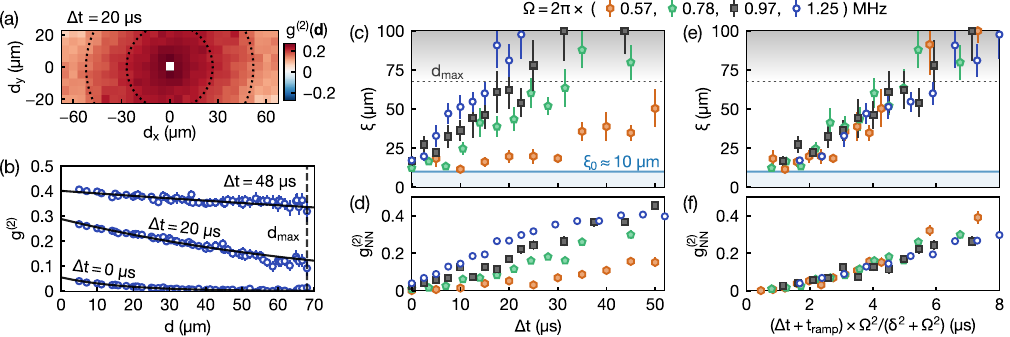}
    \caption{ 
    Dynamics of Rydberg avalanches. We consider a $5\times14$ array of $50S_{1/2}$ Rydberg-dressed atoms under cryogenic operation.
    (a) $g^{(2)}(\mathbf d)$ at $\Delta t=\SI{20}{\micro\second}$ for $\Omega=2\pi\times\SI{1.25}{\mega\hertz}$ under cryogenic operation. 
    Dashed lines indicate \SI{70}{\percent} and \SI{50}{\percent} contours.
    (b) One-dimensional $g^{(2)}(d=|\mathbf{d}|)$ at selected $\Delta t$.  
    Solid lines are fits to $A\exp(-d/\xi)$ used to extract the correlation amplitude and length. The maximum separation accessible in the array is  $d_\text{max}=\SI{68}{\micro\meter}$ (black dashed line).
    (c) Measured $\xi(\Delta t)$ for numerous Rydberg Rabi frequencies $\Omega$.
    The blue horizontal line marks $\xi_0\simeq\SI{10}{\micro\meter}$, the facilitation radius for $n=50$ (Appendix~\ref{supp:dipolar}). The exponential fits for $\xi$ become poorly constrained for $\xi>d_{\max}$, corresponding to the shaded region.
    (d) Nearest-neighbor correlation $g^{(2)}_\text{NN}=g^{(2)}(d=\SI{5}{\micro\meter})$ versus $\Delta t$.
    (e) The data in (c) are plotted versus the rescaled time $\Delta t'$ in Eq.~\eqref{eq:t_rescaled}.
    (f) The data in (d) is plotted versus $\Delta t'$. 
    }
    \label{fig:correlation}
\end{figure*}

\section{Avalanche dynamics}\label{sec:dynamics}

Our measurements reveal an additional aspect of Rydberg avalanches in atom arrays: the avalanches grow in spatial extent over time, and with increasing correlation strength. This finding extends those of Ref.~\cite{festa2022blackbody}, where the $g^{(2)}(\mbf{d})$ correlation length $\xi_0$ was proposed to arise from dipolar pair potentials. Our results suggest a dynamical correlation length $\xi(\Delta t)$ that starts at $\xi_0$ and then grows as avalanches propagate. This highlights the importance of both cryogenic suppression, to inhibit avalanche formation, and active-removal techniques~\cite{cao2025autoionization}, to stop already formed avalanches from spreading.

The spatiotemporal evolution of the avalanches is reported in Fig.~\ref{fig:correlation}. We focus on dressing with the $n=50$ state in the $5 \times 14$ array; we observe similar results for $n=60$. The correlations remain approximately isotropic in the atom plane over time, fitting well to an exponential decay profile 
\begin{equation}
g^{(2)}(d)=A(\Delta t)\exp[-d/\xi(\Delta t)],    
\end{equation}
where $d$ is the interatomic distance [Fig.~\ref{fig:correlation}(a,b)]. Both the correlation amplitude $A(\Delta t)$, which is typically well approximated by the nearest-neighbor correlation $g^{(2)}_\text{NN}$, and the correlation length $\xi(\Delta t)$ increase over time. The amplitude can surpass 0.4 after sufficient dressing time, indicating strongly correlated loss given the bound $|g^{(2)}|\leq 1$. Cryogenic operation suppresses the initial linear growth of $A(\Delta t)$ by approximately $\SI{20}{\percent}$, consistent with the cryogenic suppression of $\ava$ (Appendix~\ref{supp:g2NN}). 
$\xi(\Delta t)$ is consistent with $\xi_0$ at early dressing times but then increases, growing at a rate of approximately $\SI{3}{\meter/\second}$ for our nominal operating conditions. 
$\xi(\Delta t)$ surpasses the spatial extent of the $5\times 14$ array around $\SI{20}{\micro\second}$, after which the correlations become effectively independent of distance. 
We do not resolve whether cryogenic operation modifies the $\xi(\Delta t)$ growth rate; this will be clarified in future studies with lower $T_\text{eff}$.
In total, the data support an avalanche spreading mechanism in which contaminants facilitate Rydberg excitation in nearby dressed atoms, which locally increases the rate of contaminant generation by BBR-induced decay.

The measured avalanche propagation dynamics depends strongly on the Rydberg coupling strength $\Omega$ [Fig.~\ref{fig:correlation}(c,d)]. Both $\xi(\Delta t)$ and $A(\Delta t)$ evolve more slowly using smaller $\Omega$, and both quantities collapse well onto single curves when plotted against the rescaled time
\begin{equation}\label{eq:t_rescaled}
    \Delta t' = (\Delta t + t_\text{ramp})\frac{\Omega^2}{\delta^2+\Omega^2}, 
\end{equation}
where $t_\text{ramp}=\SI{6}{\micro\second}$ is the ramp-on and ramp-off time for the Rydberg coupling, so that $\Delta t+t_\text{ramp}$ is the effective dressing duration. This suggests that the facilitation rate scales as $\Omega^2/(\delta^2+\Omega^2)$.

Fermi's golden rule supports the facilitation rate scaling with $\Omega^2/(\delta^2+\Omega^2)$.
Facilitation occurs when a BBR-generated contaminant increases the Rydberg-dressing fraction in a neighboring atom through dipole-dipole interactions. This process is best understood in the single-atom eigenbasis of $\hat{H}$ [Eq.~\eqref{eq:dressingHam}]. Generically, there is a dark eigenstate $\ket{D} = \cos\phi\ket{g} - \sin\phi \ket{r}$ and a bright eigenstate $\ket{B} = \sin\phi\ket{g} + \cos\phi \ket{r}$ with a mixing angle $\phi=-\atan(\Omega/\delta)/2$. For small mixing angles, $\ket{D}$ has a smaller Rydberg population than $\ket{B}$. Thus, facilitation occurs when a contaminant state $\ket{P}$ induces a coupling between bright and dark states.
This coupling is provided by the dipole-dipole Hamiltonian $\hat{H}_\text{dd}$ [Eq.~\eqref{eq:Hdd}]. An initial pair state with one dressed atom and one contaminant, $\ket{D}_i \ket{P}_j$, is coupled to a bright pair state $\ket{P}_i \ket{B}_j$ with a matrix element given by
\begin{equation}
    \bra{P}_i\bra{B}_j \hat{H}_\text{dd} \ket{D}_i\ket{P}_j = -\hbar J_{ij}\cos(\phi)\sin(\phi). 
\end{equation}
The golden rule provides a simple estimate of the transition rate to bright states given the above matrix element. Higher-order effects are not expected to alter the scaling of the transition rate when there are sufficiently many atoms in the array (Appendix~\ref{supp:goldenrule}). The estimated transition rate to bright states is 
\begin{equation}
\begin{split}
    \Gamma_\text{fac}&=2\pi \rho J^2_{ij}\cos^2(\phi)\sin^2(\phi)  \\
    &= \frac{\pi}{2}\rho J^2_{ij}\frac{\Omega^2}{\delta^2+\Omega^2},  
\end{split}
\end{equation}
where $\rho$ is the spectral density of states at the final-state energy. The spin-exchange rate $J_{ij}$ has the largest energy scale and is the primary factor determining the energy bandwidth, and so $\rho$ should depend only weakly on $\Omega$ and $\delta$. Fermi's golden rule thus predicts that the facilitation rate scales as $\Omega^2/(\delta^2+\Omega^2)$, in agreement with the collapse of the measured data. 

\begin{figure}[!t]
    \centering
    \includegraphics[width=\linewidth]{ 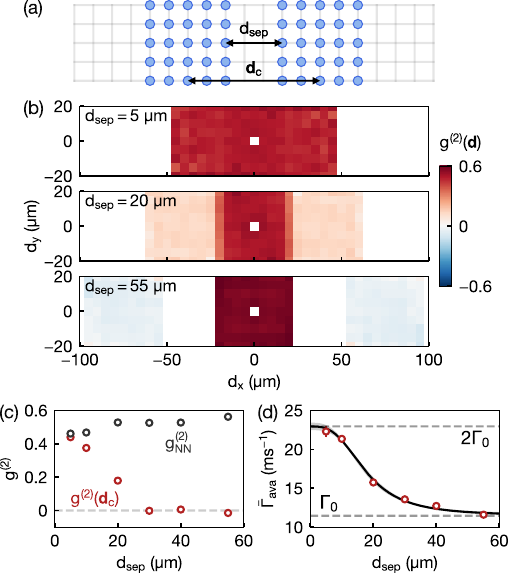}
    \caption{ 
    Decay of spatial correlations and avalanche rate in two separated $5\times 5$ subarrays.
    (a) Array geometry with lattice spacing $a=\SI{5}{\micro\meter}$, edge-to-edge subarray separation $d_{\rm sep}$, and center-to-center separation $\mathbf d_c$.
    (b) Representative spatial correlation functions $g^{(2)}(\mbf{d})$ at $d_{\rm sep}=\SI{5}{\micro\meter}$, \SI{20}{\micro\meter}, and \SI{55}{\micro\meter}. The dressing time is chosen near the point with mean survival near \SI{50}{\percent}.
    (c) Inter-array correlation $ g^{(2)}(\mathbf d_c)$ (red circles), and the 1D nearest-neighbor correlation $ g^{(2)}_{\rm NN}$ (gray circles).
    Both quantities are averaged over three dressing times with mean survival near \SI{50}{\percent} at each separation to improve statistical convergence.
    The dashed line marks zero correlation.
    (d) Mean avalanche rate $\bar{\Gamma}_{\rm ava}$ versus $d_{\rm sep}$, obtained by averaging the avalanche rates as in Eq.~\eqref{eq:ava} extracted separately for the two subarrays.
    The solid curve is a fit to the phenomenological model in Eq.~\eqref{eq:superradiance_effective_model}; the shaded band indicates fit uncertainty.
    }
    \label{fig:superradiance}
\end{figure}

While we find that avalanches dynamically grow in size, we also find that their growth can be halted by introducing spatial gaps in the atom array. To demonstrate this, we consider two $5\times 5$ subarrays that are spaced by a variable amount $d_\text{sep}$ from edge to edge [Fig.~\ref{fig:superradiance}(a)]. The full $g^{(2)}(\mbf{d})$ is shown near the $\SI{50}{\percent}$ survival point during dressing for various $d_\text{sep}$ in Fig.~\ref{fig:superradiance}(b). Strong correlations persist within each subarray, corresponding to the central $\SI{40}{\micro\meter}\times \SI{40}{\micro\meter}$ square region in $g^{(2)}(\mbf{d})$, while the cross correlations between subarrays decay to zero with increasing $d_\text{sep}$. Thus, a sufficiently large gap inhibits contaminant facilitation between subarrays. 

We quantify the decay of cross correlations by comparing the nearest-neighbor ${g}^{(2)}_\text{NN}$ to a representative cross correlation ${g}^{(2)}(\mbf{d}_\text{c})$, where $\mbf{d}_\text{c}$ is the separation vector between the centers of the subarrays [Fig.~\ref{fig:superradiance}(c)]. 
Both quantities are averaged over three dressing times with mean survival near \SI{50}{\percent} for each separation to reduce statistical fluctuations.
The ${g}^{(2)}_\text{NN}$ remains near 0.5 while ${g}^{(2)}(\mbf{d}_\text{c})$ decays to zero over roughly $\SI{30}{\micro\meter}$. 
Moreover, the average avalanche rate within each of the two subarrays, $\bar\Gamma_\text{ava}$, is found to decay to half of its original value over a similar separation scale~[Fig.~\ref{fig:superradiance}(d)]. This factor of 2 reduction is consistent with the $N$ scaling of the avalanche rate. As the subarrays split into two disconnected atom arrays, $N$ reduces from 50 to 25. Neglecting the small contribution from photoionization, we find that a simple phenomenological model based on a $1/d_\text{sep}^3$ dipolar separation dependence captures the decay of $\bar\Gamma_\text{ava}$ well. We model
\begin{equation} 
\bar{\Gamma}_{\rm ava}(d_{\rm sep}) = 
\Gamma_0
\left[
1+\frac{1}{1+(d_{\rm sep}/r_c)^3}
\right],
\label{eq:superradiance_effective_model}
\end{equation}
where $\Gamma_0$ is the avalanche rate for an isolated $5\times5$ array and $r_c$ is an effective correlation decay length. A least-squares fit yields $\Gamma_0=\SI{11.5(2)}{\per\milli\second}$ and $r_c=\SI{17.5(1)}{\micro\meter}$ with reduced chi-squared $\chi_\nu^2=0.70$. This $r_c$ is of similar magnitude as the early-time $g^{(2)}(\mbf{d})$ correlation length $\xi(\Delta t) \gtrsim \SI{10}{\micro\meter}$. In total, the data are consistent with avalanches propagating via finite-range facilitation of Rydberg excitations and contaminant generation.   

\section{Discussion and outlook}

Our results demonstrate that cryogenic environments mitigate correlated atom loss in Rydberg atom arrays.
We confirm BBR as the dominant microscopic seed for avalanche decay while identifying many factors that modify the avalanche rate: phase noise, Rydberg-Rydberg interactions, Purcell enhancement, and weakly interacting contaminants are all contributing effects captured by Eq.~\eqref{eq:ava}. Photoionization is identified as an additional avalanche trigger relevant to two-photon Rydberg coupling schemes. Avalanches are observed to propagate over time with a growth rate that scales according to Fermi's golden rule, but they can be halted by inserting extended gaps in the atom array. Taken together, our results provide a quantitative foundation for modeling avalanche rates and propagation dynamics.

The cryogenic suppression factors reported here are a proof of principle: With the key mechanisms driving avalanche decay now experimentally verified, a primary output of our work is a blueprint to achieve a much higher degree of cryogenic suppression. Of primary importance is to fully shield the atom array with materials that are not just at cryogenic temperature, but that are electrically conductive to inhibit MW radiation from passing through optical apertures. This enforces equilibration of the blackbody environment to cryogenic temperature~\cite{hassan2025cryogenic}, rather than to a mixed-temperature environment. The atom array enclosure should ideally contain cryogenic material with high MW emissivity to assist in this thermal equilibration and to mitigate Purcell effects. Employing a conductive coating on optical apertures in the current apparatus should yield equilibration to a uniform MW effective temperature $T_\text{eff}\approx\SI{45}{\kelvin}$ and this modification alone is expected to suppress $\BBR$ and $\ava$ by about $\SI{85}{\percent}$ compared to room-temperature operation. Once full cryogenic shielding is attained, previously sub-dominant trigger channels like trap photoionization become more relevant. Our work shows that dressing in shallow traps is an effective mitigation strategy, while highlighting single-photon Rydberg coupling to avoid this effect.

The BBR photons driving Rydberg decay for $n\geq 50$ have centimeter-scale wavelengths, far exceeding the spatial extent of the atom array. Thus, BBR photons may couple collectively to all atoms in the Rydberg state, and induce collective decay. This situation is analogous to that of Dicke superradiance~\cite{dicke1954coherence}, in which an ensemble of atoms all in an optically excited state collectively decays in a superradiant burst when all the atoms have sub-wavelength separations. In Rydberg atom arrays, the optical transition is replaced with a long-wavelength, BBR-induced MW transition to an opposite-parity state. A superradiant burst could then lead to all-to-all correlated loss across the atom array, similar to avalanche decay. However, the presence of multiple decay channels and strong dipolar interactions may modify or suppress the superradiant dynamics~\cite{masson2024dicke,mok2024universal}. At present, the lack of correlated loss observed between $\SI{50}{\micro\meter}$-separated $5\times 5$ atom arrays [Fig.~\ref{fig:superradiance}] provides no evidence for a superradiant burst, but it may become a formidable source of correlated loss in larger arrays. The role of cryogenic operation with regard to superradiant dynamics represents an interesting avenue for future exploration.

Our work has clear implications for error mitigation in neutral-atom quantum computing architectures. We demonstrate cryogenic lifetime enhancement of the $n\geq 50$ Rydberg $nS$ states commonly used in leading platforms. This should directly reduce error rates in entangling gates~\cite{evered2026high}.
Practical quantum processor architectures execute large numbers of entangling gates in parallel, with atoms closely packed in interaction zones and with sequential gates occurring in rapid succession~\cite{bluvstein2026fault}. 
In this context, BBR-induced contaminants could induce both erasure errors, lowering the error correction threshold, and correlated Pauli errors, reducing the effectiveness of quantum error correction~\cite{cong2022hardware} and increasing logical failure rates.
Cryogenic operation should passively reduce these errors by inhibiting BBR-driven decay to contaminant states, while active removal techniques~\cite{cao2025autoionization} could be intermittently applied between gates to eliminate residual contaminants.   

\begin{acknowledgments}
\textit{Acknowledgments:} The authors thank Ting-Wei Hsu and Matteo Marinelli for experimental assistance, and Lukas Homeier, Tim de Jongh, Christian Gross, Simon Hollerith, Ana Asenjo-Garcia, Eric Sierra, and Trey Porto for helpful discussions.  This material is based upon work supported by the U.S. Department of Energy, Office of Science National Quantum Information Science Research Centers, Quantum Systems Accelerator (Award No. DE-SCL0000121). Additional support is acknowledged from NSF QLCI award OMA-2016244 and OSI 2552714, NSF 2210527, and the Baur-SPIE Chair at JILA. B.P.M. acknowledges support from an NRC Postdoctoral Research Associateship at NIST.

\textit{Competing interests:} A.M.K. and D.H.S. are also Research Scientists at Google Quantum AI. 

\textit{Data availability:} The data supporting these findings are publicly available~\cite{avaData}.
\end{acknowledgments}

\bibliographystyle{prsty_all}
\bibliography{cryobib.bib}

\makeatletter
\def\apptoc@section{section}
\renewcommand{\addcontentsline}[3]{%
  \def\apptoc@type{#2}%
  \ifx\apptoc@type\apptoc@section
    \savedaddcontentsline{#1}{#2}{#3}%
  \fi}
\makeatother
\renewcommand{\tocname}{Appendices}
\clearpage
\tableofcontents

\appendix

\renewcommand{\theequation}{S\arabic{equation}}
\renewcommand{\thefigure}{S\arabic{figure}}
\setcounter{figure}{0}
\setcounter{equation}{0} 
\setcounter{figure}{0}
\setcounter{equation}{0}

\section{Atom preparation and Rydberg coupling}
\label{supp:atom-prep-ryd-coupling}
Atoms are loaded into static \SI{852}{\nano\meter} optical-tweezer arrays generated by a spatial light modulator (SLM) (Meadowlark Optics PDM1K-700-900-PC8). 
$\Lambda$-enhanced gray-molasses cooling on the $D_1$ line yields a single-site loading probability of up to \SI{85}{\percent}. 
For the many-body measurements, atoms are initially loaded into a $5\times20$ array with \SI{5}{\micro\meter} spacing and subsequently rearranged into defect-free target arrays using mobile \SI{852}{\nano\meter} tweezers generated by a crossed acousto-optic deflector (AOD) pair (AA Opto-Electronic DTSXY-400-850.930-002). 
The static and mobile tweezer beams are combined on a polarizing beam splitter with orthogonal linear polarizations. 
Following the respective array-preparation sequence, a \SI{3.5}{\gauss} quantization field is applied along the long axis of the array.
Finite-field $\Lambda$GM cooling yields a temperature of approximately \SI{12.0(2)}{\micro\kelvin} measured by the release-and-recapture method~\cite{tuchendler2008energy}.
The atoms are then prepared in the stretched ground state $\ket{g}=\ket{5S_{1/2},F=2,m_F=2}$ using $\sigma^+$-polarized light resonant with the $F=2\rightarrow F'=2$ transition on the $D_2$ line, together with repumping light addressing the $F=1\rightarrow F'=2$ transition.

The prepared ground state is coupled to $\ket{r}=\ket{nS_{1/2},m_J=1/2}$ through the intermediate $\ket{6P_{3/2},F'=3,m_F'=3}$ state (Fig.~\ref{fig1}). 
A $\sigma^+$-polarized \SI{420}{\nano\meter} field, red-detuned from the intermediate state by approximately \SI{1.5}{\giga\hertz}, and a $\sigma^-$-polarized counter-propagating \SI{1013}{\nano\meter} field drive the two-photon transition. 
The \SI{420}{\nano\meter} beam has an approximately circular Gaussian profile with a waist of \SI{40}{\micro\meter}, whereas the \SI{1013}{\nano\meter} beam has an elliptical profile with waists of approximately \SI{11}{\micro\meter} and \SI{140}{\micro\meter}; its narrow axis is aligned with the axial direction of the optical tweezers.

\section{Rydberg lifetime measurement procedure}
\label{supp:lifetime-measurement}

\begin{figure}[t!]
    \centering
    \includegraphics[]{ 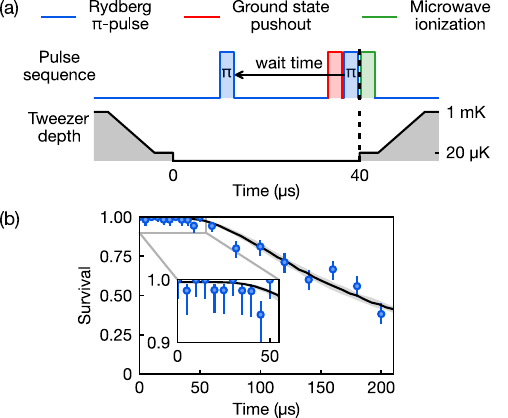}
    \caption{Single-atom Rydberg-lifetime sequence and release-and-recapture characterization.
    (a) Rydberg lifetime measurement sequence. 
    The tweezer depth is reduced from \SI{1}{\milli\kelvin} to \SI{20}{\micro\kelvin} before the traps are extinguished for a fixed \SI{40}{\micro\second} interval. 
    Simultaneous \SI{420}{\nano\meter} and \SI{1013}{\nano\meter} pulses (blue) drive two Rydberg $\pi$ pulses separated by a variable wait time; both fields are extinguished during the wait. 
    A ground-state pushout pulse (red) immediately precedes the second $\pi$ pulse, and MW ionization (green) immediately follows it.
    (b) Release-and-recapture survival versus trap-off time for $\mathrm{NA}=0.230(2)$ at trap depth of \SI{20}{\micro\kelvin}. 
    The solid black curve is a release-and-recapture fit yielding an atomic temperature of \SI{0.93(6)}{\micro\kelvin}, and the shaded region denotes the fit uncertainty. 
    The inset shows the first \SI{50}{\micro\second} on an expanded vertical scale.
    }
    \label{fig:fig_sup_rydberg_seq_and_RR}
\end{figure}

The Rydberg lifetime measurement sequence is shown in Fig.~\ref{fig:fig_sup_rydberg_seq_and_RR}(a).
For the single-atom Rydberg-lifetime measurements, atoms are loaded stochastically into a dilute $1\times4$ array with \SI{33}{\micro\meter} spacing; no rearrangement is performed.
This large separation suppresses resonant dipole-exchange interactions between atoms transferred to nearby $P$ contaminant states and atoms remaining in the target $nS$ state.
Following cooling and state preparation, the tweezer depth is linearly reduced from \SI{1}{\milli\kelvin} to \SI{20}{\micro\kelvin} over \SI{10}{\milli\second} to lower the atom temperature, after which the traps are extinguished. 
For these measurements, the tweezer beam incident on the objective is apertured using the SLM to reduce the effective numerical aperture to $\mathrm{NA}=0.230(2)$, thereby increasing the trap waist and recapture volume. 
At the final depth, release-and-recapture thermometry using classical Monte Carlo simulations of a Gaussian trapping potential  yields an atomic temperature of \SI{0.93(6)}{\micro\kelvin}. 
Together, the reduced numerical aperture and lower temperature provide recapture probabilities above \SI{98}{\percent} for measured trap-off durations up to \SI{40}{\micro\second} [Fig.~\ref{fig:fig_sup_rydberg_seq_and_RR}(b)].

After the tweezers are extinguished, a resonant two-photon $\pi$ pulse transfers the atom from $\ket{g}$ to $\ket{r}$. 
Both Rydberg-coupling fields are then extinguished during a variable wait time and subsequently reapplied to drive a second $\pi$ pulse, which returns population remaining in $\ket{r}$ to $\ket{g}$. 
The two-photon Rabi frequency is $\Omega/2\pi\approx\SI{2.5}{\mega\hertz}$, with approximately matched single-photon Rabi frequencies to minimize scattering from the $6P_{3/2}$ state during the pulse.
Extinguishing the \SI{1013}{\nano\meter} field during the wait time is necessary to avoid coupling $\ket{r}$ to the $6P_{3/2}$ state, which introduces scattering and would bias the measured lifetime toward lower values.

Two removal steps ensure that atoms that decay from the target Rydberg state during the wait time are not subsequently counted as surviving $\ket{r}$ population. 
Without these steps, decay products that remain trappable could generate a wait-time-dependent survival offset and bias the measured lifetime towards higher values.
First, radiative decay can populate trappable states within the $5S_{1/2}$ hyperfine manifold. Atoms decaying to $\ket{g}$ are resonantly excited back to $\ket{r}$ by the second $\pi$ pulse and subsequently removed via MW ionization, as explained below. However, radiative decay also populates other states in the $5S_{1/2}$ manifold which are not addressed by the second $\pi$ pulse.
These decayed, ground-state atoms are removed immediately before the second $\pi$ pulse using a few-microsecond pulse with a single-pass pushout beam resonant with the $F=2\rightarrow F'=3$ transition, together with repumping light addressing the $F=1\rightarrow F'=2$ transition.
Second, BBR-induced transitions populate nearby contaminant Rydberg states that are not addressed by the second $\pi$ pulse and may subsequently decay into trappable ground states. 
These contaminants are removed immediately after the second $\pi$ pulse using a \SI{9}{\micro\second} MW-ionization pulse applied through an antenna inside the cold box~\cite{zhang2025lifetime}. 
Although the applied pulse lasts \SI{9}{\micro\second}, ionization typically occurs in less than \SI{1}{\micro\second}, minimizing the probability that contaminants decay into trappable states during removal.
The pushout and MW-ionization steps also suppress residual signals associated with imperfect $\pi$ pulses, such that pulse infidelity primarily affects the overall detection amplitude rather than producing a finite survival offset.
For every wait time, the MW-ionization pulse directly follows the second $\pi$ pulse, while the total trap-off interval is held fixed at \SI{40}{\micro\second} to eliminate wait-time-dependent release-and-recapture loss. 
The tweezers are then restored and ramped back to \SI{1}{\milli\kelvin} before imaging.

The lifetime measurement sequence was independently characterized for every principal quantum number studied, with residual false-survival probabilities below \SI{0.1}{\percent}. 
The measured survival probabilities are fit to zero-offset exponential decay curves versus wait time, and the resulting $1/e$ time constants are reported as the Rydberg-state lifetimes [Fig.~\ref{fig1}(d) and Fig.~\ref{fig2:direct_lifetime}].

\section{Cold box microwave environment}\label{supp:COMSOL}

The boundary conditions defined by the cold box can modify the MW mode density at the position of the atoms, thus suppressing or enhancing the decay rate of Rydberg states via the Purcell effect. We estimate the Purcell-modified decay rates
using finite-element simulation in COMSOL as follows. 

Ref.~\cite{pultinevicius2025long} considered a similar problem, wherein the Purcell factor inside a parallel plate capacitor was estimated through finite-element simulations. The authors constructed a small dipole antenna inside the capacitor plates, and measured the antenna impedance while driving an AC current to emit a MW radiation field. By relating the impedance inside the capacitor to that of free space, the authors estimate the Purcell factor.

We take a different approach than Ref.~\cite{pultinevicius2025long} that we find is less sensitive to the exact antenna geometry.  
We instead compute the power $P$ radiated into the cold box by a point dipole, which is related to the Purcell factor by $F_P=P/P_0$ where $P_0$ is the total power radiated into free space~\cite{antennapurcell}. $P$ is calculated by first solving for the steady-state MW field induced by a point dipole at the location of the atoms, then integrating the Poynting vector through a small sphere surrounding the dipole. The radius of the sphere does not affect the integrated power due to Poynting's theorem.  
We choose a radius of $\sim\SI{0.5}{\mm}$, which is large enough to avoid numerical errors due to the diverging electric field very close to the dipole and small enough to avoid intersecting the cold box itself. An absorbing, perfectly matched layer is placed around the simulated volume with a thickness and size large enough to avoid nonphysical reflections due to mesh discretization. $P_0$ is calculated in the same simulated volume for consistency, but simply without the cold box present.

\begin{figure}[t!]
    \centering
    \includegraphics[width=\linewidth]{ 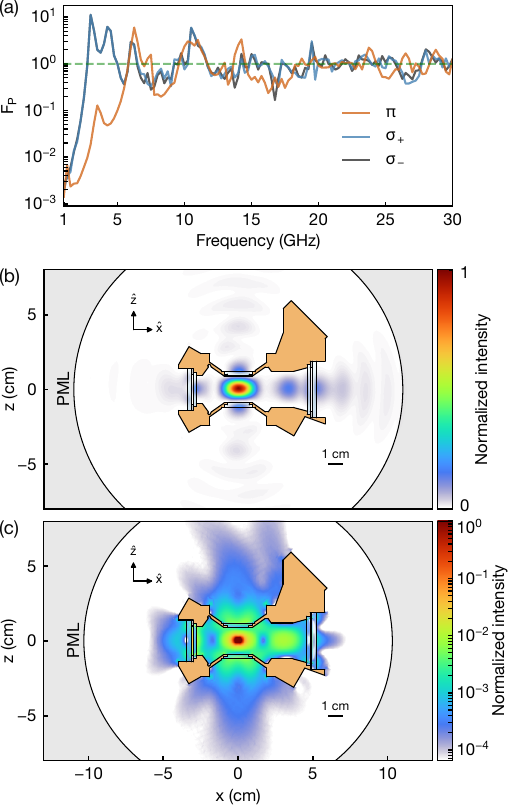}
    \caption{Finite-element simulations for the Purcell factor $F_P$. (a) Simulated Purcell factors inside the cold box versus frequency and MW polarization. 
    (b) Example eigenmode identified at \SI{10.83}{\GHz} with a $Q$-factor of approximately 32. The eigenmode is calculated using an iterative Krylov solver. 
    The outer black circle marks the beginning of the perfectly matched layer, which extends into the gray region.  
    (c) Example steady-state emission pattern for a radiating point dipole, linearly polarized along $\hat{x}$ and driven at the $70S_{1/2}\rightarrow70P_{3/2}$ transition frequency ($\sim\SI{10.65}{\GHz}$). The cold box modifies the emission pattern compared to the free-space dipole emission pattern.
    }
    \label{fig:comsol}
\end{figure}

The computed Purcell factors are plotted in Fig.~\ref{fig:comsol}(a). The asymmetry of the cold box yields different Purcell factors for all three polarizations. These Purcell factors are used for calculating the Purcell-modified lifetimes described by Eq.~\eqref{eq:BBR} for room-temperature measurement and shown in Fig.~\ref{fig2:direct_lifetime}(a). The Purcell factors are computed at the exact theoretical transition frequencies for each $nS$ state. The Einstein $A$ coefficients and Rydberg transition frequencies are calculated using the Alkali.ne Rydberg Calculator (ARC)~\cite{vsibalic2017arc}.
Our simulations extend to \SI{30}{\giga\hertz}. This limits our estimates of the Purcell factor to only those transition frequencies corresponding to the dominant $nS\to nP$ and $nS\to(n-1)P$ transitions, which account for approximately \SI{70}{\percent} of the total BBR decay rates. The Purcell factors for transition frequencies exceeding those we can simulate are assumed to be unity, which is supported by the trend of the Purcell factors approaching unity with increasing MW frequency in Fig.~\ref{fig:comsol}(a).

We compare the measured decay rates under cryogenic operation using a simplified model with the same $\BBR$ formula in Eq.~\eqref{eq:BBR} but using a weighted average of the number of thermal MW photons $\bar n(\omega,T)$ at $T=\SI{300}{\kelvin}$ and $T=\SI{45}{\kelvin}$. Specifically, we use \begin{equation}
    \bar n_\text{Cryo}(\omega) = f \bar n(\omega,\SI{45}{\kelvin}) + (1-f)\bar n(\omega,\SI{300}{\kelvin})
\end{equation} 
The fraction $f$ is an effective BBR-shielding factor for the cold box. As reported in the main text, fitting the Rydberg lifetimes for $n\leq 60$ yields $f=\SI{44(3)}{\percent}$, giving an estimate of the percentage of room-temperature BBR that is suppressed by the cold box. The effective temperature is calculated by fitting the BBR decay rates in Fig.~\ref{fig2:direct_lifetime}(b) with $n\leq60$ to the form of $\BBR$ given in Eq.~\eqref{eq:BBR} with the temperature as a fit parameter, yielding $T_{\rm eff}=\SI{188(8)}{\kelvin}$. 
This provides a simple benchmark for the overall suppression of BBR-induced decay.

The geometry of the cold box used in simulations is a simplified model that does not include smaller conductive components (e.g. the electrodes) or external structures (e.g. the objective lens or vacuum chamber). However, calculations of more complete geometric models do not significantly modify the extracted decay rates, and we therefore use the simplified model, which permits simulations at higher frequencies than would otherwise be possible.

\section{Rydberg decay induced by laboratory\\microwave noise}\label{supp:MWnoise}

\begin{figure}[ht]
    \centering
    \includegraphics[width=\linewidth]{ 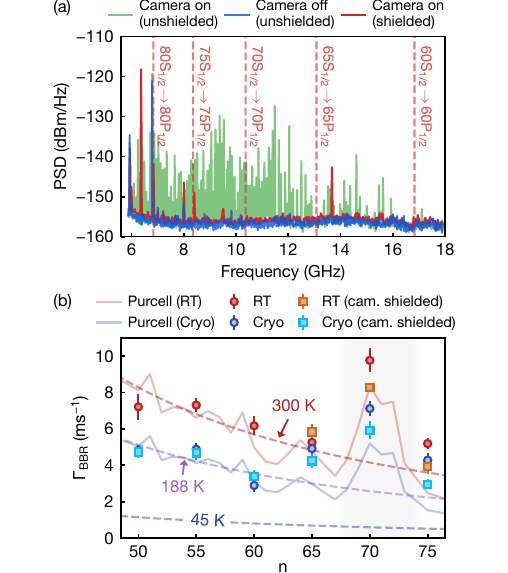}
    \caption{
    Suppression of CMOS camera-generated MW radiation.
    (a) Input-referred MW power spectral density measured adjacent to the fluorescence-imaging camera with the camera on and unshielded (green), off and unshielded (blue), and on and shielded (red). 
    The vertical dashed lines mark the $nS_{1/2}\rightarrow nP_{1/2}$ transition frequencies for $n=60$, 65, 70, 75, and 80.
    (b) Microwave-induced decay rate as a function of principal quantum number under room-temperature and cryogenic conditions, measured with unshielded (circles) and shielded (squares) camera. 
    Shielded-camera measurements cover $n=50$--75 under cryogenic conditions and $n=65$, 70, and 75 at room temperature.
    The MW-induced decay rate is extracted as in Fig.~\ref{fig2:direct_lifetime}, where the solid and dashed curves and the shaded region are also defined.
    }
    \label{fig:fig_sup_shielded_camera_lifetime}
\end{figure}

Our atom-imaging camera, a scientific complementary metal-oxide-semiconductor (CMOS) camera, is found to emit broadband MW radiation. Camera-generated radiation is found to reduce Rydberg lifetimes and enhance contaminant-state production by triggering the same decay channels as BBR.   
To suppress this noise, we surround the camera with a $\SI{2}{\centi\meter}$ layer of MW-absorbing material (ECCOSORB AN-75) and enclose the entire structure within a thin copper shield, leaving a hole for the camera aperture.

The camera's MW emission spectrum is measured over \SIrange{5.9}{18}{\giga\hertz} using a broadband antenna (RF SPIN QRH20E) connected to a MW amplifier (Mini-Circuits ZVE-3W-183+). We measure under three conditions: camera off, camera on without shielding, and camera on with shielding.
The power spectral densities are normalized by the frequency-dependent amplifier gain but still retain noise added by the amplifier. Consequently, the camera-off spectrum lies above the room-temperature Johnson--Nyquist noise level $k_BT\simeq-174~\mathrm{dBm/Hz}$ for a matched \SI{50}{\ohm} load.
With the unshielded camera, the excess MW noise reaches approximately \SI{30}{\decibel} above the Johnson--Nyquist level and extends across frequencies resonant with transitions from $nS$ to $n'P_J$ Rydberg states for $n>60$ [Fig.~\ref{fig:fig_sup_shielded_camera_lifetime}(a)].

Our camera shield reduces the camera-on MW spectrum to the camera-off background throughout the measured frequency range.
Consistent with this spectral suppression, the measured cryogenic Rydberg lifetimes increase for $n>60$, with a maximum reduction of \SI{31(9)}{\percent} in the MW-induced decay rate for $n=75$ [Fig.~\ref{fig:fig_sup_shielded_camera_lifetime}(b)]. 
For $n\leq60$, no statistically significant change in lifetime is observed upon adding the shielding. The $n>60$ lifetime data reported in the main text are acquired with the camera shielded, whereas the $n\leq60$ lifetime and Rydberg-dressing data are acquired without shielding.

The lifetime of the $80S_{1/2}$ state is found to be \SI{26(2)}{\micro\second}, substantially shorter than the predicted room-temperature lifetime of \SI{204}{\micro\second}. 
Neither shielding of the camera nor cryogenic operation produces a statistically significant difference in the measured lifetime. The reduced lifetime is most likely due to the $80S\to80P$ transition frequency being close to the $^{87}$Rb ground-state hyperfine splitting near \SI{6.8}{\giga\hertz}. This frequency is routinely used in the laboratory and cannot be fully extinguished during Rydberg experiments. Thus, it is likely that ambient MW noise near this frequency increases the decay rate of the $80S_{1/2}$ state. We omit $n=80$ from the data presented in the main text for this reason.
Consistent with this interpretation, a prominent signal in the laboratory MW noise spectrum is observed near this frequency [Fig.~\ref{fig:fig_sup_shielded_camera_lifetime}(a)].
Additional MW isolation, for example through electrically conductive coatings~\cite{wang2025can} on the cold windows, could suppress this effect in future designs.  

\section{Calculation of dipolar facilitation range}\label{supp:dipolar}

The calculation of avalanche correlation lengths using dipolar pair potentials was introduced in Ref.~\cite{festa2022blackbody}. We follow the procedure described therein to estimate the predicted correlation lengths $\xi_0$ for the $50S_{1/2}$ and $60S_{1/2}$ states in \Rb. In summary, the potentials shown in Fig.~\ref{fig3:avalanche} describe the energy spectrum for pair eigenstates of the dipole-dipole Hamiltonian. Atom one is considered to be in a desired $nS_{1/2}$ state while atom two is conditioned on being in one of the contaminant states $P_i$ as a result of a BBR-induced transition, represented by the state $\ket{S,P_i}$.
The dipolar interaction between the $S$-state atom and contaminant atom modifies the pair spectrum. This can bring the otherwise off-resonant dressing laser, detuned by $\delta$ from the bare ground-Rydberg transition, into resonance with a pair eigenstate. The dressing laser can therefore resonantly excite interacting pair eigenstates and accelerates loss; this is the facilitation mechanism.
The dipolar interaction depends on the particular $\ket{P_i}$ state populated by the BBR transition. Thus, we calculate the pair potentials separately for each contaminant state.

For each contaminant state $\ket{P_i}$, the two-atom dipole-dipole Hamiltonian is diagonalized at every interatomic separation $R$ to yield eigenstates $\ket{\psi_{\alpha i}(R)}$ with energies $E_{\alpha i}(R)$, where $\alpha$ indexes the eigenstates. The calculations take the internuclear axis to be parallel to the quantization axis. The blackbody photon producing the contaminant state supplies the energy difference for the $S\to P_i$ transition, leaving an interaction-induced energy difference $\Delta E_{\alpha i}(R)=E_{\alpha i}(R)-E_{\alpha i}(R=\infty)$~\cite{festa2022blackbody}. Resonant facilitation occurs when $\Delta E_{\alpha i}=\hbar \delta$. The $R$ at which this condition is met determines $\xi_0$, as shown in Fig.~\ref{fig3:avalanche}. For $\delta=-2\pi\times\SI{2}{\mega\hertz}$ used in this work, $\xi_0\simeq\SI{13}{\micro\meter}$ for $n=60$ and $\xi_0\simeq\SI{10}{\micro\meter}$ for $n=50$. 
Because several pair eigenstates and contaminant states contribute, $\xi_0$ represents an approximate facilitation range rather than a uniquely defined radius. 

The color intensity in Fig.~\ref{fig3:avalanche} shows the overlap $w_{\alpha i}(R)=|\braket{\psi_{\alpha i}(R)}{S,P_i}|^2$.
This overlap determines the coupling rate to each pair eigenstate following BBR-induced decay.
Given a contaminant state $\ket{P_i}$, the dressing laser couples $\ket{g,P_i}$ to $\ket{\psi_{\alpha i}(R)}$ with an effective Rabi frequency $\Omega_{\alpha i}(R)=\Omega\braket{\psi_{\alpha i}(R)}{S,P_i}$ and hence $w_{\alpha i}(R)=|\Omega_{\alpha i}(R)/\Omega|^2$.
Thus, pair eigenstates are visible in Fig.~\ref{fig3:avalanche} only when dipolar interactions admix $\ket{S,P_i}$ character; the other pair eigenstates are not coupled by the dressing laser.
Darker shading indicates pair eigenstates that are more strongly coupled, and thus more relevant to avalanche dynamics.
Unlike the shading used in Ref.~\cite{festa2022blackbody}, the color does not include the BBR transition rate into $\ket{P_i}$; our shading describes the facilitated excitation rate conditional on a contaminant already being present.

\section{Rydberg dressing and avalanche measurements}\label{supp:DressParams}

\subsection{Dressing conditions and scattering}
For the avalanche measurements, the rearranged arrays described in Sec.~\ref{supp:atom-prep-ryd-coupling} are initialized using the same finite-field $\Lambda$GM cooling and state-preparation sequence as in the single-atom measurements. The tweezers are operated at their full effective numerical aperture, $\mathrm{NA}>0.48$, instead of the reduced numerical aperture used for single-atom lifetime measurements.
The tweezer depth is reduced before dressing from \SI{0.88}{\milli\kelvin} to \SI{8.8}{\micro\kelvin}.
The lower intensity reduces inhomogeneous tweezer light shifts and, more importantly, suppresses photoionization of residual $6P_{3/2}$ population by the \SI{852}{\nano\meter} tweezer light.

Under the nominal dressing conditions, the single-photon Rabi frequencies are $\Omega_{1013}/2\pi\simeq\SI{150}{\mega\hertz}$ and $\Omega_{420}/2\pi\simeq\SI{25}{\mega\hertz}$, yielding a two-photon Rabi frequency of $\Omega/2\pi\simeq\SI{1.25}{\mega\hertz}$. 
This coupling ratio is chosen to minimize occupation of the intermediate $6P_{3/2}$ state, thereby suppressing spontaneous scattering and tweezer-induced photoionization.
Further details of the photoionization channel are given in Sec.~\ref{sec:ionization}.
In the Rabi-frequency scans shown in Fig.~\ref{fig:avaStar}(a) and Fig.~\ref{fig:correlation}, $\Omega$ is varied by changing only $\Omega_{420}$, while $\Omega_{1013}$ is held fixed.

The \SI{420}{\nano\meter}-induced scattering rate through $6P_{3/2}$ is $\Gamma_{\mathrm{sc}}^{420}\simeq(\Omega_{420}/2\Delta)^2\Gamma_{6P}=\SI{575}{\per\second}$ in the weak-dressing approximation.
Accounting for the $6P_{3/2}$ branching fraction into the $F=1$ manifold, approximately $\SI{15.6}{\percent}$, gives a predicted depumping rate of $\Gamma_{\mathrm{dep}}^{(420)}\simeq\SI{90}{\per\second}$, in reasonable agreement with the measured value of \SI{84(2)}{\per\second}.
The \SI{1013}{\nano\meter} field introduces an additional scattering channel by coupling the Rydberg component of the dressed state back to $6P_{3/2}$. 
Despite the small Rydberg admixture, the strong \SI{1013}{\nano\meter} coupling makes this the dominant contribution, with $\Gamma_{\mathrm{sc}}^{1013}/\Gamma_{\mathrm{sc}}^{420}\simeq3.5$. 
The combined intermediate-state scattering rate is therefore $\Gamma_{\mathrm{sc}}^{6P}\simeq\SI{2.6}{\per\milli\second}$. 

In addition to intermediate-state scattering, spontaneous decay from the Rydberg state occurs during dressing at rates of approximately \SI{0.58}{\per\milli\second} for $n=50$ and \SI{0.33}{\per\milli\second} for $n=60$ at the nominal Rydberg dressing strength. 
Weak optical-pumping and repumping fields are applied throughout the dressing pulse to return population that reaches the ground-state hyperfine manifold to $\ket{g}$. 
This recycling allows atoms to re-enter the dressing cycle following spontaneous emission and approximately preserves the number of atoms participating in the dressing dynamics.

\subsection{Independent loss characterization}

Distinguishing avalanche decay from independent, single-atom loss requires characterizing the independent loss rate. The measured independent loss rates are used to compute the $\SI{95}{\percent}$ confidence intervals, as shown in Fig.~\ref{fig3:avalanche}(d), that we use to distinguish avalanche events from independent atom loss. The independent-loss rate is measured in dressing experiments using a single atom, rather than a full atom array, but using otherwise matched dressing parameters. Although optical pumping restores the atoms' internal state following spontaneous emission, it does not reverse the recoil heating accumulated during scattering. Thus, loss occurs due to recoil heating in addition to BBR-induced loss. This means that the single-atom loss rate cannot be identified directly with $P_{\rm Ryd}\Gamma_{\rm BBR}$. 
Similar contributions are expected from recoil heating and BBR-induced loss, with their sum roughly consistent with the measured single-atom loss rate.

For the phase-noise and tweezer-induced ionization scans (Fig.~\ref{fig:systematic}), the independent-loss rate may vary with the scanned parameter. For that data, the independent-loss rate is instead extracted directly from the same many-atom survival distribution used to determine the avalanche rate.
Avalanche events produce a low-survival tail below the expected binomial distribution arising from independent atom loss. Thus, the independent-loss component is inferred from the high-survival portion of the survival distribution. To do so, we use a survival threshold $f_{\rm th}$ corresponding to the $40$th percentile of the survival distribution, meaning that $\SI{60}{\percent}$ of shots with the highest survival fraction are considered. This is an effective method to filter out avalanche events for short dressing times $\Delta t$. The high-survival shots are used in a maximum likelihood estimate to find the independent-loss probability $p$ at a given $\Delta t$, and we then fit $p(\Delta t)$ to an exponential decay to extract the independent-loss rate. 

The maximum likelihood fit works as follows. For a fixed $\Delta t$, we label $M_k$ as the number of loaded atoms in shot $k$ and $S_k$ as the number of surviving atoms after dressing. The conditional likelihood function for a given independent-loss probability $p$ is
\begin{equation}
\mathcal{L}(p)
=
\prod_{k}
\frac{
P_{\rm binom}(S_k\mid M_k,1-p)
}{
P_{\rm binom}(S_k/M_k\geq f_{\rm th}\mid M_k,1-p)
},
\end{equation}
where $P_{\rm binom}(S_k\mid M_k,1-p)$ is the binomial probability for $S_k$ successes given $M_k$ trials with probability $1-p$. $P_{\rm binom}(S_k/M_k\geq f_{\rm th}\mid M_k,1-p)$ is the binomial probability of having a survival fraction of at least $f_{\rm th}$. Normalizing by this factor accounts for truncating the data and avoids an upward bias that would result from fitting the selected shots with an ordinary binomial likelihood. Maximizing $\mathcal{L}(p)$ gives the maximum likelihood estimate for $p$ at the given $\Delta t$.

The measured independent-loss rate does not account for the suppression of $P_{\rm Ryd}$ due to van der Waals interactions in the dressed array and may therefore differ from the true independent-loss rate in the array. However, this effect does not significantly influence the classification of avalanche shots. Avalanche events tend to take out large fractions of the atom array. Small variations in the independent-loss confidence interval have little effect on identifying avalanche events, consistent with the sensitivity analysis discussed in Appendix~\ref{sec:robust}. 

Note that single-atom loss events do not necessarily imply an equivalent decrease in the number of atoms participating in the avalanche dynamics. An atom that escapes its tweezer trap does not immediately fly away, and can still participate in facilitation dynamics during Rydberg dressing. 
Thus, the effective number of atoms available to seed or propagate an avalanche can remain approximately unchanged even as the final detected survival decreases. This provides a natural explanation for the approximately constant avalanche-onset rate observed versus dressing time [Fig.~\ref{fig3:avalanche}(e)] despite the gradual independent loss of atoms.

\section{Avalanche data analysis}\label{supp:AvaDataAna}

\subsection{Robustness of avalanche rate extraction}\label{sec:robust}
Avalanche events are identified by comparing the measured survival distribution with the binomial distribution expected for independent single-atom loss. 
A shot is classified as an avalanche when its surviving-atom number falls below the $\SI{95}{\percent}$ confidence interval for independent loss.
To quantify the dependence on this threshold, we repeat the rate extraction analysis over a broad range of confidence intervals.

\begin{figure}[t]
    \centering
    \includegraphics[width=\linewidth]{ 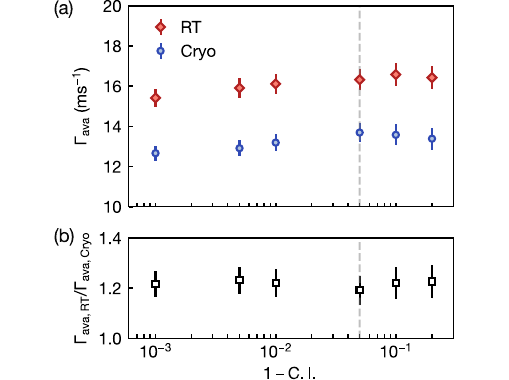}
    \caption{Extracted avalanche rate versus the confidence interval used to classify correlated loss, shown for a $5\times14$ array with $n=60$ dressing.
    (a) $\Gamma_{\rm ava}$ at room temperature (red diamonds) and under cryogenic operation (blue circles) as a function of $1-\mathrm{C.I.}$
    (b) The ratio of $\Gamma_{\rm ava}$ at room temperature and under cryogenic operation.
    The vertical dashed line marks the \SI{95}{\percent} confidence interval used for results in the main text.
    }
    \label{fig:fig_sup_avarate_robustness}
\end{figure}

Figure~\ref{fig:fig_sup_avarate_robustness} shows this analysis for the $5\times14$ array with $60S_{1/2}$ dressing. 
The absolute avalanche rates show only minor variation over more than two orders of magnitude in the confidence interval. Moreover, the ratio of avalanche rates between room-temperature and cryogenic operation remains constant to within measurement uncertainty.
Similar results are observed for all array sizes and for $n=50$: the extracted rates and ratios typically vary by less than \SI{10}{\percent} over more than two orders of magnitude variation in the confidence interval.
The confidence interval therefore introduces only a small systematic uncertainty and does not affect the observed scaling with atom number and Rydberg admixture, or the consistency with $\Gamma_{\rm ava}^{\star}\simeq N P_{\rm Ryd}\Gamma_{\rm BBR}^{\rm trig.}$.

In addition, our analysis permits up to \SI{4}{\percent} defect fraction after rearrangement. 
The avalanche rates extracted from the strictly defect-free subset of data and the data with up to $\SI{4}{\percent}$ defect fraction agree to within their fit uncertainties.

\subsection{Evaluation of $g^{(2)}(\mathbf{d})$}

Here we show how the $g^{(2)}(\mbf{d})$ correlation function is computed in the general case where there may be defects in the rearranged atom array. We use the same survival variables as in Eq.~\eqref{eq:g2} of the main text, where $s^{(k)}_i=+1$ indicates atom survival for shot $k$, else $s^{(k)}_i=-1$ indicating loss. 
We now account for loading defects. Let $f_i^{(k)}=1$ if site $i$ is occupied in shot $k$, else $f_i^{(k)}=0$ indicating a defect.
The mean $s_i$ at each site is
\begin{equation}
\bar{s}_i=
\frac{\sum_k^\text{shots} f_i^{(k)}s_i^{(k)}}
{\sum_k^\text{shots} f_i^{(k)}},
\end{equation}
where the sum is taken over all experimental shots.
The correlation function is then evaluated as
\begin{equation}
g^{(2)}(\mathbf d)=
\frac{
\displaystyle
\sum_k^\text{shots}\sum_{\substack{i,j \\ \mathbf r_i-\mathbf r_j=\mathbf d}}
f_i^{(k)}f_j^{(k)}
\left(s_i^{(k)}-\bar{s}_i\right)
\left(s_j^{(k)}-\bar{s}_j\right)
}{
\displaystyle
\sum_k^\text{shots}\sum_{\substack{i,j \\ \mathbf r_i-\mathbf r_j=\mathbf d}}
f_i^{(k)}f_j^{(k)}
},
\label{eq:g2_supp}
\end{equation}
where the sum over $i,j$ is taken over all atom pairs whose interatomic separation vector satisfies $\mathbf r_i-\mathbf r_j=\mathbf d$ for a given $\mathbf d$.
Site-resolved means $\bar{s}_i$ are subtracted before averaging over site pairs and experimental shots. This is the correct method to compute correlations for the general case where the $\bar s_i$ may vary. 
Only pairs for which both sites are occupied contribute, and every eligible pair in every shot is weighted equally. 
For defect-free arrays, Eq.~\eqref{eq:g2_supp} reduces to Eq.~\eqref{eq:g2} of the main text. 
The radial correlation $g^{(2)}(d)$ is evaluated using the same estimator, but with the sums in both the numerator and denominator using the condition $|\mathbf r_i-\mathbf r_j|=|\mathbf d|=d$.

\section{Additional avalanche triggers and\\systematic effects}\label{supp:avaTrigs}

\subsection{Interaction-suppressed dressing fraction}
\label{sec:dressingfraction}

Van der Waals interactions between Rydberg-dressed atoms can suppress simultaneous Rydberg excitation in nearby atoms~\cite{johnson2010interactions,pupillo2010strongly}. This reduces the average Rydberg dressing fraction relative to the single-atom value and generates anticorrelations between neighboring Rydberg excitations. 
This is the same interaction blockade mechanism that gives rise to the soft-core form of the effective interaction between Rydberg-dressed atoms. In our context, the reduced Rydberg fraction results in a slower BBR-induced decay rate, and this effect should be taken into account to achieve agreement with the predicted rate in Eq.~\eqref{eq:ava}.

To account for this effect, the per-atom Rydberg fraction
$P_{\rm Ryd}=\frac{1}{N}\sum_i\langle \hat n_i\rangle$
is calculated by exact diagonalization of Eq.~\eqref{eq:dressingHam} for a $5\times5$ square lattice with periodic boundary conditions, using the experimental lattice spacing, detuning, and van der Waals coefficient. The results are reported in Fig.~\ref{fig:systematic}(d).
The calculation is performed using QuSpin~\cite{weinberg2019quspin}. 
Periodic boundary conditions eliminate edge effects and restore translational symmetry, allowing the calculation to be restricted to the zero-momentum sector and thereby reducing the Hilbert-space dimension, which makes exact diagonalization of the $25$-site system tractable.

Convergence toward the bulk limit is assessed by comparison with calculations using open boundary conditions. 
The Rydberg fraction at interior sites of increasingly large open arrays approaches the periodic-boundary result, while the reduced interaction blockade near the boundaries leads to enhanced Rydberg populations.
For the $50S_{1/2}$ and $60S_{1/2}$ states, the Rydberg fractions at non-corner edge sites exceed the periodic-boundary values by approximately \SI{5}{\percent} and \SI{16}{\percent}, respectively. 
Extrapolating these enhancements to the experimental $5\times14$ array gives estimated corrections to the array-averaged Rydberg fraction of approximately \SI{2}{\percent} and \SI{7}{\percent}, respectively, neglecting corner contributions. 
Including the enhanced populations at the corners increases these estimates to approximately \SI{3}{\percent} and \SI{9}{\percent}. 
We use the periodic-boundary result as a simple estimate of the bulk Rydberg fraction relevant to the experimental arrays. 

\subsection{Weakly interacting contaminant states} 
\label{sec:seedingefficiency}
\begin{figure}[t]
    \centering
    \includegraphics[]{ 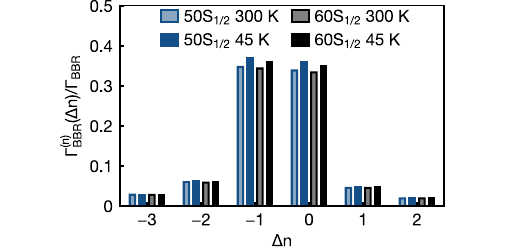}
    \caption{
    BBR branching fractions into contaminant manifolds.
    Calculated fractional transition rates from $nS_{1/2}$ to $(n+\Delta n)P_J$ for $n=50,60$ under isotropic \SI{300}{\kelvin} and \SI{45}{\kelvin} conditions.
    Each bar sums over $J=1/2,3/2$ and the final-state magnetic sublevels and is normalized by the total BBR-induced transition rate for the corresponding initial state and thermal conditions.
    }
    \label{fig:relative_bbr_vs_n}
\end{figure}

Some contaminant states have weak dipolar interactions with the $nS_{1/2}$ states and do not trigger avalanches. We estimate the fraction of BBR-induced decay events at room temperature that do trigger avalanches as follows. We compute the dipolar interaction strength for all relevant contaminant states and check whether each is greater in magnitude than the dressing laser detuning $\delta$. This is a simple estimate to determine whether the contaminant can trigger resonant facilitation. We then compute the probability of decaying to a sufficiently strongly interacting contaminant state to estimate an avalanche seeding fraction given a BBR-induced decay event.

Dipolar interactions between $nS_{1/2}$ and contaminant $(n+\Delta n)P_J$ states are evaluated as $V_{\rm dd}=C_3/R^3$ at the nearest-neighbor separation $R=\SI{5}{\micro\meter}$. 
The $C_3$ coefficients are calculated using the PairInteraction package~\cite{mogerle2026accurate} at a bias field of \SI{3.5}{\gauss}, for $n=50,60$ and $J=1/2,3/2$. 
The mean strength $\overline{V}_{\rm dd}$ is obtained by averaging over the four nearest-neighbor interactions, two along the bias-field axis and two perpendicular to it. Comparing $\overline{V}_{\rm dd}$ to the dressing-detuning magnitude $|\delta|/2\pi=\SI{2}{\mega\hertz}$ indicates that the $(n-1)P$ and $nP$ manifolds are the only contaminant channels that are estimated to be strongly interacting enough to trigger avalanches [Fig.~\ref{fig:systematic}(e)]. 
The sum of BBR-induced transition rates into these two manifolds defines $\BBR^\text{trig.}$. 

\begin{table}[t]
\centering
\caption{BBR branching fractions $f_n$ into the $(n-1)P$ and $nP$ manifolds in thermal equilibrium at $\SI{300}{\kelvin}$ and $\SI{45}{\kelvin}$. Calculated using ARC~\cite{vsibalic2017arc}. 
}
\setlength{\tabcolsep}{10pt}
\begin{tabular}{ccc}
\hline
Initial state & $f_n$ ($\SI{300}{\kelvin}$) & $f_n$ ($\SI{45}{\kelvin}$) \\
\hline
$50S_{1/2}$ & $0.686$ & $0.729$ \\
$60S_{1/2}$ & $0.677$ & $0.711$ \\
\hline
\end{tabular}
\label{tab:bbr_branching}
\end{table}

Let $\Gamma_{\rm BBR}^{(n)}(\Delta n)$ denote the BBR-induced transition rate from $nS_{1/2}$ to the $(n+\Delta n)P$ manifold, summed over the final-state fine-structure and magnetic sublevels. 
Figure~\ref{fig:relative_bbr_vs_n} shows these rates normalized by the total BBR-induced transition rate at both $\SI{300}{\kelvin}$ and $\SI{45}{\kelvin}$.
The fraction of BBR-induced decay events corresponding to the strongly interacting $nP$ and $(n-1)P$ states is
$f_n =\left[\Gamma_{\rm BBR}^{(n)}(-1)+\Gamma_{\rm BBR}^{(n)}(0)\right]/\BBR$, where $\BBR$ is the corresponding total BBR-induced decay rate.
The calculated values are listed in Table~\ref{tab:bbr_branching}. 
Their mean is approximately $0.70$, motivating the estimate $\BBR^\text{trig.}\approx0.7\BBR$ used in the main text for room temperature. Under cryogenic operation, non-uniform thermal equilibration across MW frequencies prevents there from being an easily calculated ratio between $\BBR$ and $\BBR^\text{trig.}$. 
Purcell enhancement is omitted from these transition-rate calculations; Purcell effects do not significantly affect the $n=50$ and $n=60$ decay rates~[Fig.~\ref{fig2:direct_lifetime}].

\subsection{Correction for drift in Rabi frequency}

The Rydberg Rabi frequency and resonance frequency are calibrated for each experimental setting. 
For measurements at fixed dressing strength, the nominal parameters are $\Omega_0/2\pi=\SI{1.25}{\mega\hertz}$ and $\delta_0/2\pi=-\SI{2}{\mega\hertz}$. 
Repeated resonance calibrations show negligible variations in $\delta_0$.
However, slow drift in the measured Rabi frequency $\Omega$ results in a relative standard deviation of up to $\SI{3}{\percent}$ between experiments.
Fluctuation in $\Omega$ systematically shifts the avalanche rates by modifying $\PRyd$. To correct for this in Fig.~\ref{fig3:avalanche}(h), each measured avalanche rate is normalized by the ratio of $\PRyd$ calculated using the measured $\Omega$ and the nominal $\PRyd$ calculated using $\Omega_0$.
This approximate normalization follows the BBR-seeding term in Eq.~\eqref{eq:ava} and accounts for the interaction-induced suppression of $P_{\rm Ryd}$.

The estimated ionization contribution to $\ava$ is also rescaled with the same correction factor derived from the total Rydberg population $\PRyd$.
This rescaling is adopted for simplicity rather than being derived from the microscopic ionization process, because a change in $\PRyd$ does not uniquely determine the corresponding change in the ionization rate.
Its impact is expected to be small because ionization contributes only a small fraction of the total rate [Fig.~\ref{fig:systematic}(b)].
We separately consider the phase-noise contribution, whose expected $\Omega^2$ dependence is known~\cite{krinner2024low}, while retaining the $\PRyd$ scaling for all other systematic corrections.
Relative to scaling the full avalanche rate by $\PRyd$ alone, this treatment changes the fitted avalanche-rate slopes by less than \SI{1}{\percent}, well below their reported uncertainties.
We therefore use the simpler $\PRyd$ normalization throughout the supplementary analysis and for the quantitative comparison in Fig.~\ref{fig:avaStar}(b).
All other measured avalanche rates in the main text are presented without any normalization.

\subsection{Rydberg laser phase noise}
\label{sec:phasenoise}

\begin{figure}[!t]
    \centering
    \includegraphics[]{ 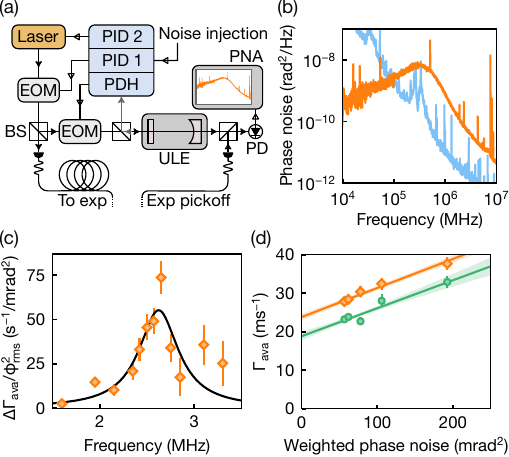}
    \caption{
    Laser phase noise and its contribution to $\ava$.
    (a) Stabilization and phase-noise measurement of the \SI{1013}{\nano\meter} laser. 
    Pound--Drever--Hall (PDH) feedback acts on the laser current and piezo using proportional-integral-derivative (PID) controller~2. PDH feedback also acts on a fast fiber electro-optic modulator (EOM) through PID~1~\cite{endo2018residual}. Noise is injected through the setpoint of PID~1. 
    The beat note between the light transmitted through the ultralow-expansion (ULE) cavity and light sampled after two amplification stages is analyzed with a phase noise analyzer (PNA).
    (b) Phase-noise power spectral densities of the \SI{1013}{\nano\meter} (orange) and \SI{420}{\nano\meter} (blue) fields. 
    The latter is inferred from measurements at \SI{840}{\nano\meter} by multiplying the power spectral density by four to account for frequency doubling.
    (c) Avalanche-rate response $\Delta\Gamma_{\rm ava}/\phi_{\rm rms}^2$ versus single-tone modulation frequency for a room-temperature $5\times14$ array dressed with the $50S_{1/2}$ state. 
    Here, $\Delta\Gamma_{\rm ava}$ is the rate increase relative to zero added noise, and $\phi_{\rm rms}^2$ is the measured laser phase variance of the injected tone. 
    The black curve is a Lorentzian fit constrained to zero vertical offset.
    (d) $\ava$ versus response-weighted phase noise for $5\times14$ arrays dressed with $50S_{1/2}$ under room-temperature (green circles) and cryogenic (orange diamonds) operation. 
    The noise level is varied by white-noise injection. 
    The weighted variance is obtained by integrating the sum of the \SI{1013}{\nano\meter} and \SI{420}{\nano\meter} phase-noise power spectral densities against the fitted Lorentzian profile in (c), normalized to unit peak.
    Lines show linear fits; shaded bands indicate fit uncertainty.
    }
    \label{fig:phasenoise}
\end{figure}

Laser phase noise adds a small spectral component to the Rydberg dressing laser that is resonant with the Rydberg transition. This leads to incoherent excitation of the Rydberg state beyond the expected dressing fraction, increasing the dressed-state BBR-induced decay rate and thereby increasing the avalanche rate~\cite{festa2022blackbody}.
The \SI{1013}{\nano\meter} phase noise is measured from the beat note between light that transmits through a cavity frequency reference and
light sampled after two amplification stages, just before delivery to the atoms [Fig.~\ref{fig:phasenoise}(a)], following the method of Ref.~\cite{schmid2019simple}.
The phase noise for the \SI{840}{\nano\meter} light, which is frequency doubled to $\SI{420}{\nano\meter}$ for two-photon Rydberg excitation, is measured using an optical phase-noise analyzer (OEwaves). 
The \SI{420}{\nano\meter} phase-noise spectrum is inferred by multiplying the \SI{840}{\nano\meter} spectral density by four to account for frequency doubling.
The \SI{1013}{\nano\meter} field dominates the phase noise in the relevant frequency range [Fig.~\ref{fig:phasenoise}(b)].

We characterize the effects of phase noise by injecting signals into the setpoint of a fast fiber-EOM feedback loop for the \SI{1013}{\nano\meter} laser.
The loop bandwidth of a few megahertz covers the frequencies near $\delta$, the range relevant to Rydberg dressing.
The avalanche response to single-tone phase modulation is characterized by scanning the modulation frequency in a room-temperature $5\times14$ array dressed with the $50S_{1/2}$ Rydberg state at fixed dressing conditions.
For each modulation frequency, the change in avalanche rate $\Delta\Gamma_{\rm ava}$ is evaluated relative to the rate without phase modulation. 
The phase variance $\phi_{\rm rms}^{2}$ of the injected tone is obtained by integrating its excess phase-noise spectrum after background subtraction. 
The resulting response $\Delta\Gamma_{\rm ava}/\phi_{\rm rms}^{2}$ is fitted by a Lorentzian constrained to zero offset [Fig.~\ref{fig:phasenoise}(c)]. The susceptibility has a bandwidth of about a megahertz, most likely set by interaction-induced broadening of the Rydberg resonance in the $5\times 14$ dressed array. 

The response to broadband phase noise is calibrated by varying the amplitude of white noise injected into the \SI{1013}{\nano\meter} feedback loop. 
For each noise level, the \SI{1013}{\nano\meter} and \SI{420}{\nano\meter} phase-noise power spectral densities are summed, assuming uncorrelated phase fluctuations between the two fields. 
This sum is integrated against the measured single-tone response function normalized to unit peak amplitude~[Fig.~\ref{fig:phasenoise}(c)]. 
The same normalized response function is used for both room-temperature and cryogenic measurements.
The avalanche rate increases approximately linearly with the resulting weighted phase variance over the measured range at both temperatures [Fig.~\ref{fig:phasenoise}(d)].
We therefore determine the absolute phase-noise susceptibility directly from the slope of this broadband calibration, using the single-tone measurement only to specify the relative frequency weighting.
For comparison, using the absolute amplitude of the fitted Lorentzian from the room-temperature single-tone scan predicts a smaller avalanche-rate increase than is measured under broadband noise injection.
Deviations at the high-frequency end of the single-tone scan, together with possible response outside the scanned frequency range, may contribute to this discrepancy.
Importantly, this discrepancy does not enter the extraction of the phase-noise contribution to the avalanche rate, which is based on the directly measured broadband response rather than on the absolute amplitude of the single-tone Lorentzian fit.

The observed linearity of $\ava$ versus phase noise supports an empirical calibration over the measured noise range. 
Using this calibration, the combined phase-noise contribution to the avalanche rate, $\Gamma_{\rm PN}$, is estimated as the difference between $\ava$ under nominal operating conditions (without any added noise) and the $\ava$ extrapolated to zero total weighted phase variance.
In the simplified avalanche-onset model from Eq.~\eqref{eq:ava}, this contribution corresponds to $\Gamma_{\rm PN}=NP_{\rm PN}\BBR^\text{trig.}$.
For $5\times14$ arrays ($N=70$) dressed with the $50S_{1/2}$ state, this gives $\Gamma_{\rm PN}=\SI{4.1(9)}{\per\milli\second}$ under cryogenic operation and \SI{4.3(4)}{\per\milli\second} at room temperature, corresponding to \SI{18(3)}{\percent} and \SI{15(1)}{\percent} of the respective avalanche rates. 
This extrapolation assumes that the response remains linear down to zero noise, as expected in the small-signal limit, and that the frequency dependence of the response is approximately unchanged between the two temperatures.
Within the simple picture in which phase noise creates additional Rydberg population $P_{\rm PN}$ that subsequently seeds avalanches through BBR-induced decay, changing the BBR environment is expected primarily to rescale the response amplitude through $\BBR^{\rm trig.}$ rather than change the shape of the spectral response.
Response amplitude rescaling changes the fitted broadband susceptibility but has no effect on the inferred $\Gamma_{\rm PN}$.

The extracted phase-noise contributions to $\ava$ for $50S_{1/2}$ are similar at the two temperatures, with
$\Gamma_{\rm PN}^{\rm Cryo}/\Gamma_{\rm PN}^{\rm RT}=0.95(23)$.
For comparison, the independently measured BBR-induced decay rates give
$\Gamma_{\rm BBR}^{\rm Cryo}/\Gamma_{\rm BBR}^{\rm RT}=0.67(8)$.
These ratios are statistically consistent.
Because an independent phase-noise calibration was not performed for $60S_{1/2}$, we estimate its correction by rescaling the $50S_{1/2}$ result using the measured total $\BBR$ for the relevant Rydberg state and temperature (Fig.~\ref{fig:avaStar}(b)). 
This approximation may differ from the microscopic triggering-rate scaling when the cryogenic branching fraction depends on frequency, but the effect on the final comparison is limited because phase noise is a subdominant contribution to $\ava$.

\subsection{Tweezer-induced photoionization}
\label{sec:ionization}
\begin{figure}
    \centering
    \includegraphics[]{ 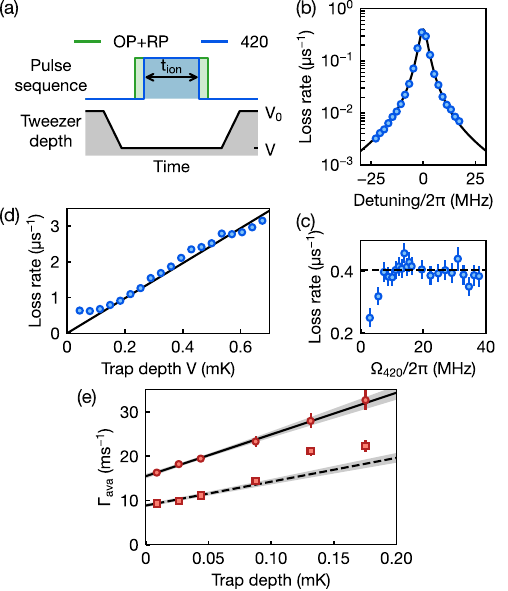}
    \caption{Tweezer-induced photoionization and its contribution to avalanche decay.
    (a) Sequence for single-atom photoionization measurements.
    The \SI{852}{\nano\meter} tweezer depth is ramped from $V_0$ to $V$ before applying a \SI{420}{\nano\meter} pulse of variable duration $t_{\rm ion}$.
    Optical pumping (OP) and repumping (RP) are applied before and during this pulse to recycle population into $\ket{5S_{1/2},F=2,m_F=2}$.
    The loss rate is extracted from the survival probability versus $t_{\rm ion}$.
    (b) Single-atom loss rate versus \SI{420}{\nano\meter} detuning from the $|5S_{1/2},F=2,m_F=2\rangle \rightarrow |6P_{3/2},F'=3,m_F'=3\rangle$ transition, at $V=\SI{0.1}{\milli\kelvin}$ and $\Omega_{420}/2\pi=\SI{2.9(5)}{\mega\hertz}$.
    The solid curve is a Lorentzian fit.
    (c) Loss rate versus resonant \SI{420}{\nano\meter} Rabi frequency.
    The dashed line marks the saturated loss rate.
    (d) Loss rate versus tweezer depth $V$.
    The solid line shows a calculation based on a model Rb atomic potential, with no adjustable parameters.
    (e) Avalanche rate versus tweezer depth for $5\times14$ (circles) and $5\times8$ (squares) arrays dressed to $60S_{1/2}$ at room temperature.
    The solid line is a linear fit to the $5\times14$ data.
    The dashed line uses the fitted slope scaled by the atom-number ratio $40/70$ and passes through the lowest-depth $5\times8$ data point.
    Shaded bands indicate fit uncertainty.
    }
    \label{fig:ionization}
\end{figure}

Ionization is an established tool for detecting and controlling Rydberg populations in neutral-atom experiments. 
In alkaline-earth atoms, excitation of the ionic core can produce doubly excited states above the ionization threshold that decay by autoionization~\cite{cooke1978doubly,xu1986sr,xu1994rydberg}. 
This process enables efficient Rydberg-state detection~\cite{lochead2013number,mcquillen2013imaging,madjarov2020high} and rapid removal of contaminant Rydberg atoms~\cite{cao2025autoionization}. 
Charges generated by photoionization have additionally been used to suppress stray electric fields in neutral-atom platforms~\cite{chen2026photo,deng2026efficient}.

Photoionization can also occur when the combined energies of photons absorbed from Rydberg excitation light and other optical fields exceed the ionization threshold.
In the present experiment, the combined energy of one \SI{420}{\nano\meter} photon and one \SI{852}{\nano\meter} tweezer photon exceeds the Rb ionization threshold, allowing photoionization through the intermediate $6P_{3/2}$ state. 
The resulting ions can Stark shift neighboring Rydberg levels toward resonance with the dressing laser and thereby seed correlated loss.

This channel is characterized through single-atom loss measurements using the sequence in Fig.~\ref{fig:ionization}(a). 
The $\sigma^+$-polarized \SI{420}{\nano\meter} light drives the $\ket{5S_{1/2},F=2,m_F=2}\rightarrow\ket{6P_{3/2},F'=3,m_F'=3}$ transition, while optical pumping and repumping return population redistributed by spontaneous decay to $\ket{5S_{1/2},F=2,m_F=2}$.
Loss rates are extracted by fitting the survival probability versus excitation time $t_{\rm ion}$ to an exponential decay.
The tweezer depth $V$ is calibrated from the differential light shift of the $5S$--$5P$ transition measured by pushout spectroscopy, using polarizabilities calculated with ARC~\cite{vsibalic2017arc}. 
At $V=\SI{0.1}{\milli\kelvin}$ and $\Omega_{420}/2\pi=\SI{2.9(5)}{\mega\hertz}$, the loss spectrum is well described by a Lorentzian centered at the light-shifted resonance [Fig.~\ref{fig:ionization}(b)]. 
On resonance, the loss rate saturates with increasing Rabi frequency, consistent with saturation of the $6P_{3/2}$ population [Fig.~\ref{fig:ionization}(c)].

The photoionization cross section is calculated using Fermi's golden rule. 
Energy-normalized scattering states in the continuum are obtained from a model potential for Rb~\cite{marinescu1994dispersion}, and their dipole matrix elements with the bound $\ket{6P_{3/2},F'=3,m_F'=3}$ state are evaluated. 
The predicted loss rate is the product of this cross section, the tweezer photon flux, and the $6P_{3/2}$ population fraction. 
At saturated excitation, an approximately constant excited-state population therefore gives a loss rate proportional to $V$. 
At larger trap depths, the measured loss rates agree with the parameter-free theoretical prediction of \SI{9.8}{\per\micro\second\per\milli\kelvin} [Fig.~\ref{fig:ionization}(d)].
The plateau in the measured loss rate at shallow trap depths may reflect additional loss from recoil heating.
For application to off-resonant Rydberg dressing, the model retains ionization through the populated $6P_{3/2}$ state and neglects ionization pathways through other intermediate states.

The contribution to avalanche decay is calibrated separately by varying the tweezer depth for room-temperature arrays dressed to $60S_{1/2}$ [Fig.~\ref{fig:ionization}(e)]. 
The $5\times14$ data exhibit an approximately linear dependence. 
At shallow depths, the $5\times8$ data are consistent with the fitted slope scaled by the atom-number ratio $40/70$. 
At larger depths, the $5\times8$ rates exceed this comparison; the origin of this departure is not resolved by the present measurements.

The $5\times14$ data are fitted to $\Gamma_{\rm ava}^{\rm fit}(V)=\Gamma_{\rm bg}+bV$, where $\Gamma_{\rm bg}$ is the extrapolated zero-depth rate. 
Assuming that other avalanche-seeding contributions remain approximately independent of trap depth over the fitted range, the photoionization contribution to avalanche rate at the nominal operating depth is $bV_{\rm op}$, modeled as $NP_{\rm E}\Gamma_{\rm ion}$ in Eq.~\eqref{eq:ava}.
At $V_{\rm op}=\SI{8.8}{\micro\kelvin}$, this yields $bV_{\rm op}=\SI{0.83(8)}{\per\milli\second}$, corresponding to \SI{5.1(6)}{\percent} of the fitted avalanche rate. 
For smaller arrays under otherwise matched conditions, the ionization contribution is estimated from the $5\times14$ calibration by assuming linear scaling with atom number.

An independent consistency check is obtained by estimating the first-ion production rate in the array. 
Neglecting ground–Rydberg coherence, the intermediate-state population per atom is approximated as $P_{\rm E} \simeq (1-P_{\rm Ryd})(\Omega_{420}/2\Delta)^2 + P_{\rm Ryd}(\Omega_{1013}/2\Delta)^2$. 
The many-body Rydberg fraction $P_{\rm Ryd}$ is used to include interaction-induced suppression of Rydberg excitation. 
Multiplying $NP_{\rm E}$ by the calibrated per-atom photoionization rate $\Gamma_{\rm ion}$ in $6P_{3/2}$ yields a first-ion production rate of approximately $\SI{1.1}{\per\milli\second}$ for the $5\times14$ array dressed to $60S_{1/2}$ under nominal operating conditions, comparable to the independently inferred trap-dependent avalanche contribution of $\SI{0.83(8)}{\per\milli\second}$.

\subsection{Systematic corrections for variable Rabi frequencies}
\label{sec:rabiscanCorrection}

To obtain the systematics-corrected avalanche rates in Fig.~\ref{fig:avaStar}(a), we subtract the estimated phase-noise and photoionization contributions at each Rydberg Rabi frequency $\Omega$. The measurements are performed with a $5\times14$ array dressed with $50S_{1/2}$ under cryogenic operation. During the scan, $\Omega$ is varied through the \SI{420}{\nano\meter} coupling strength while keeping the \SI{1013}{\nano\meter} coupling strength and two-photon detuning fixed, as detailed in Appendix~\ref{supp:DressParams}.

We extrapolate the calibrated phase-noise contribution at the nominal Rabi frequency $\Omega_0=2\pi\times\SI{1.25}{\mega\hertz}$ according to
\begin{equation}
\Gamma_{\rm PN}(\Omega)=N P_{\rm PN}(\Omega_0) \Gamma_{\rm BBR}^{\rm trig.}
\left(\frac{\Omega}{\Omega_0}\right)^2,
\end{equation}
where $P_{\rm PN}(\Omega_0)$ is the phase-noise-induced Rydberg population measured under nominal experiment conditions (Appendix~\ref{sec:phasenoise}).
The quadratic scaling follows from the phase-noise-induced incoherent Rydberg excitation rate: for a fixed phase-noise spectrum, the noise-induced coupling amplitude is proportional to $\Omega$, giving an excitation rate proportional to $\Omega^2$ in the weak-noise regime~\cite{krinner2024low}.
The photoionization contribution is estimated by scaling the calibration with the intermediate-state population,
\begin{equation}
\Gamma_{\rm ava}^{\rm ion}(\Omega)=
\Gamma_{\rm ava}^{\rm ion}(\Omega_0)
\frac{P_{\rm E}(\Omega)}{P_{\rm E}(\Omega_0)},
\end{equation}
where $\Gamma_{\rm ava}^{\rm ion}(\Omega_0)=N P_{\rm E}(\Omega_0) \Gamma_{\rm ion}$ is estimated from the tweezer-depth measurements with $60S_{1/2}$ dressing under otherwise comparable conditions. We approximate the excited-state population as a function of the Rydberg Rabi frequency $P_{\rm E}(\Omega)$ using the expression in Appendix~\ref{sec:ionization}, including the interaction-suppressed Rydberg population $\PRyd(\Omega)$.

The corrected avalanche rate is then
\begin{equation}
\ava^\star(\Omega)=\ava(\Omega)
-\Gamma_{\rm PN}(\Omega)
-\Gamma_{\rm ava}^{\rm ion}(\Omega),
\end{equation}
where we obtain $\PRyd(\Omega)$ by interpolating the exact-diagonalization results described in Appendix~\ref{sec:dressingfraction}, resulting in Eq.~\eqref{eq:avaCorr} in the main text when $\Omega = \Omega_0$.

\begin{figure}[!t]
    \centering
    \includegraphics[]{ 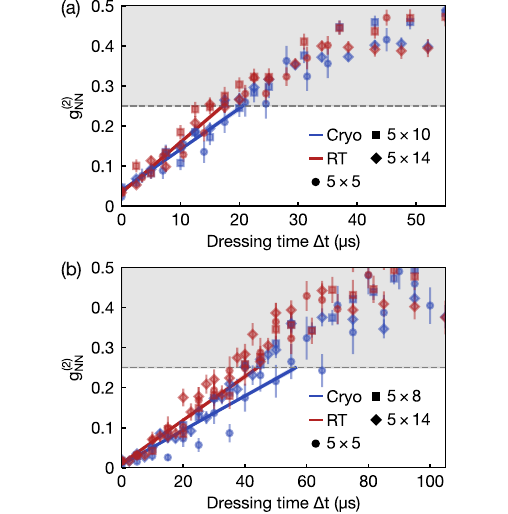}
    \caption{Cryogenic suppression of $g^{(2)}$.
    Nearest-neighbor correlations $g_{\rm NN}^{(2)}=g^{(2)}(d=\SI{5}{\micro\meter})$ versus dressing time $\Delta t$ are shown for Rydberg-dressed arrays at room temperature (red) and under cryogenic operation (blue).
    (a) $50S_{1/2}$ dressing in $5\times5$, $5\times10$, and $5\times14$ arrays.
    Linear fits to the early-time growth yield slopes of \SI{10.3(6)}{\per\milli\second} (cryo) and \SI{12.6(5)}{\per\milli\second} (RT), corresponding to a suppression of \SI{19(6)}{\percent}.
    (b) $60S_{1/2}$ dressing in $5\times5$, $5\times8$, and $5\times14$ arrays. 
    The fitted slopes are \SI{4.3(1)}{\per\milli\second} (cryo) and \SI{5.4(2)}{\per\milli\second} (RT), corresponding to a suppression of \SI{21(4)}{\percent}.
    The cryogenic suppression in $g^{(2)}$ at early times is consistent with the cryogenic suppression in $\ava$.
    }
\label{fig:fig_sup_nearest_neighbour_vs_time_total}
\end{figure}

\section{Temperature and array-size dependence of nearest-neighbor correlations}\label{supp:g2NN}

To complement the correlation function dynamics presented in Fig.~\ref{fig:correlation}, Fig.~\ref{fig:fig_sup_nearest_neighbour_vs_time_total} compares the one-dimensional nearest-neighbor correlation $g_{\rm NN}^{(2)}$ across array sizes, Rydberg states, and temperatures. 
For all configurations, $g_{\rm NN}^{(2)}$ increases with dressing time, reflecting the development of correlated loss. Cryogenic operation reduces $g_{\rm NN}^{(2)}$ at early times by an amount consistent with the cryogenic suppression in $\ava$, for both $n=50$ and $n=60$.

\section{Fermi's golden rule regime of validity}
\label{supp:goldenrule}

The argument based on Fermi's golden rule presented in the main text is an approximation.
For a finite closed system, the many-body spectrum is discrete and the transition probability is quadratic at sufficiently short times, rather than described by a constant irreversible rate.  
However, the number of dressed-state configurations accessible to a contaminant grows rapidly with system size, producing a dense spectrum of final states.
Over times that are long compared with the inverse of the bandwidth of the many-body spectrum, but short compared with recurrence times associated with the discrete spectrum, the set of final states forms a continuous spectrum to a good approximation.
Moreover, atomic motion, spatial inhomogeneity of Rydberg coupling and trap locations, and dissipative processes provide additional dephasing that further suppresses any coherent revivals. 
We therefore use the golden rule as an approximate and simple method to describe the observed scaling of the facilitation dynamics.

The derivation in the main text neglects van der Waals interactions between dressed atoms. 
These interactions suppress the many-body Rydberg fraction and can modify both the relevant matrix elements and the density of accessible final states. 
A full treatment would therefore require diagonalizing the interacting dressed many-body Hamiltonian [Eq.~\eqref{eq:dressingHam}] together with the dipole-dipole Hamiltonian [Eq.~\eqref{eq:Hdd}].
In the main text, van der Waals interactions are ignored, so that the single-atom dressed eigenbasis can be used to calculate transition rates.
The observed collapse of the correlation length and correlation amplitude data as a function of the rescaled time $\Delta t'$  predicted by the golden rule suggests that these interaction corrections do not substantially change the scaling of the avalanche growth rate.

\end{document}